\documentclass[12pt,preprint]{aastex}

\usepackage{bm}

\usepackage{graphicx}

\def\unit #1{\,{\rm #1}}
\def\mpc{\unit{Mpc}}
\def\gyr{\unit{Gyr}}
\def\hubbleunit{\unit{km\,sec^{-1}\,Mpc^{-1}}}
\def\kpc{\unit{kpc}}
\def\arcmsq{\unit{arcmin^2}}
\def\kms{\unit{km\,s^{-1}}}

\begin {document}
\slugcomment{}
\shorttitle{ }
\shortauthors{ }

%\title{Tracing the evolution in the merger-rate of galaxies \\ out to redshift $\sim1.2$}

\title{Towards a robust estimate of the merger rate evolution \\ using near-IR photometry}

\author{A. Rawat,\altaffilmark{1,2}
Francois Hammer,\altaffilmark{2}
Ajit K. Kembhavi,\altaffilmark{1}
Hector Flores,\altaffilmark{2}
}

\altaffiltext{1}{Inter University Centre for Astronomy and
Astrophysics, 
Post Bag 4, Ganeshkhind, Pune 411007, India: rawat@iucaa.ernet.in}
\altaffiltext{2}{GEPI, l'Observatoire de Paris-Meudon, 92195 Meudon, France}

\begin{abstract}
We use a combination of deep, high angular resolution imaging data from the CDFS (HST/ACS GOODS survey) and ground based near-IR $K_s$ images to derive the evolution of the galaxy major merger rate in the redshift range $0.2 \leq z \leq 1.2$. We select galaxies on the sole basis of their J-band rest-frame, absolute magnitude, which is a good tracer of the stellar mass. We find steep evolution with redshift, with the merger rate $\propto (1+z)^{3.43\pm0.49}$ for optically selected pairs, and $\propto (1+z)^{2.18\pm0.18}$ for pairs selected in the near-IR. Our result is unlikely to be affected by luminosity evolution which is relatively modest when using rest-frame J band selection. The apparently more rapid evolution that we find in the visible is likely caused by biases relating to incompleteness and spatial resolution affecting the ground based near IR photometry, underestimating pair counts at higher redshifts in the near-IR. The major merger rate was $\sim$5.6 times higher at $z\sim1.2$ than at the current epoch. Overall $41\%$$\times$($0.5\gyr$/$\tau$)  of all galaxies with $M_J\leq-19.5$ have undergone a major merger in the last $\sim8 \gyr$, where $\tau$ is the merger timescale. Interestingly, we find no effect on the derived major merger rate due to the presence of the large scale structure at $z=0.735$ in the CDFS.
%\textbf{We do find some evidence for increased star formation due to possible interactions between members of a pair.} 

%	
\end{abstract}


\section{Introduction}
%\labsecn{intro}
Galaxy mergers are believed to be the chief mechanism driving galaxy evolution within the hierarchical framework. Although mergers are rare at the current epoch, the hierarchical framework predicts that the merger rate must have been higher at earlier epochs. 
Despite its importance in understanding galaxy evolution, 
the quantification of the galaxy major merger rate and its evolution with redshift is still an ill constrained and hotly debated issue. Patton et al.~\cite{patton1997} have derived a clear increase in the merger fraction with redshift till $z\sim0.33$ with a power law index of $2.8\pm0.9$. However, extension of this work to higher redshifts has been riddled with controversy. Le Fevre et al.~\cite{lefevre2000} reported steep evolution of major merger rate till $z\sim1.0$, with a power law index of $3.2\pm0.6$ using pair-counting in the optical band for identifying the merger candidates. Bundy et al.~\cite{bundy2004} reported a much more modest evolution of merger rate using K' band images for identifying major merger candidates. Lin et al.~\cite{lin2004} also reported very weak evolution in the merger rates using the DEEP2 redshift survey. Bell et al.~\cite{bell2006} reported a fairly rapid evolution in merger fraction of massive galaxies between $z\sim0.8$ and the current epoch, using the technique of projected 2-point correlation function. Lotz et al.~\cite{lotz2006}, on the other hand, reported that the major merger fraction remains roughly constant at $\sim7\%\pm2\%$ till $z\sim1.2$, using a nonparametric technique for quantifying galaxy morphology. Some of the discrepancies in the results quoted above may be attributed to differences in sample selection criteria, different techniques to derive the merger rate and different definitions of major mergers used by various people.
%

%Some of the discrepancies in the results quoted above can be attributed to different definitions of major mergers used by different people, which effectively probes a different stage in the merger process of a galaxy pair. 
We have used a combination of high resolution imaging data from the HST/ACS GOODS survey and $K_s$ imaging data from the VLT followup of the GOODS-South field to quantify the major merger rate of galaxies and its redshift evolution using the technique of pair counting. %We discuss the possible biases that are introduced when neighbours are identified in just one of the two filters. 
Section~\ref{data} lists the datasets we have used in this work, and briefly explains the methodology that we have employed in this paper for deriving the merger rate. Section~\ref{sampleselection} details the sample selection criterion employed by us and the possible biases in our sample. Section~\ref{ks} explains the details of the photometry performed by us in the $K_s$ filter. Section~\ref{pairs} explains in detail how we identify major pairs of galaxies, along with correcting for possible contamination and incompleteness issues. Section~\ref{bias} compares the differences in pairs identified in the visible and the Near-IR. Section~\ref{evolution} deals with the calculation of major merger rate from our identified pairs and it's evolution with redshift. We conclude by discussing the implications of our results obtained in this paper in Section~\ref{discuss}.
We adopt a cosmology with $H_0$\,=$70\hubbleunit$,  $\Omega_{\rm  M}$\,=\,0.3 and $\Omega_\Lambda$\,=\,0.7. 

\section {The Data}
\label{data}
We have used version v1.0 of the reduced, calibrated images of the Chandra Deep Field South (CDFS) acquired with HST/ACS as part of the GOODS survey (Giavalisco et al. 2004). The SExtractor (Bertin and Arnouts 1996) based version r1.1 of the ACS multi-band source catalogs was used for object identification.
Spectroscopic redshifts were taken from the redshift catalog of the VVDS (Le Fevre et al. 2004), GOODS/FORS2 redshift survey (Vanzella et al. 2005, 2006) \& from the IMAGES survey (Ravikumar et al. 2007).
The near-IR J \& $K_s$ band imaging data of the GOODS/CDFS region from the ESO GOODS/EIS Release Version 1.5 (Vandame et al., in preparation) was used.  
%We also used the MIPS 24$\mu$m source catalog made available as part of the GOODS, Spitzer Legacy Data Products, Third Data Release (DR3, Dickinson et al. 2006).
%

\subsection{The methodology}
\label{methodology}
The methodology that we employ in this paper is quite straightforward. First, we identify sources with known spectroscopic redshifts in the GOODS-S field. We then shortlist those galaxies with redshift in the range $0.2 \leq z \leq 1.2$ and rest frame absolute magnitude brighter than $M_J(AB)=-19.5$ to obtain a volume limited sample of primary galaxies. We then identify neighbours within a projected radius of $5h_{100}^{-1}\kpc\leq r \leq 20h_{100}^{-1}$\kpc~ around each primary galaxy. If the difference in the apparent magnitude $\delta m$ of the primary galaxy and the neighbour satisfy the condition $-1.5 \leq \delta m \leq 1.5$ in either the $z$ or the $K_s$ filter, the galaxy pair is designated to be a {\it{major pair}} in that particular filter. This is explained in Section~\ref{pairs}. 

We do not have the redshift for the secondary member of a major pair in most cases, and as such some of them can simply be foreground/background superpositions. Such a contamination is statistically corrected for by using number counts of objects in the field in the z or $K_s$ filter and is explained in detail in Section~\ref{contamination}. Our primary galaxy sample is then binned into three redshift bins. The fraction of galaxies undergoing major merger in any given redshift bin is then inferred from the fraction of galaxies existing in major pairs in the same bin, after applying a correction for foreground/background contamination. The evolution in the major merger fraction is then determined by looking for any change in the major merger fraction over the three redshift bins that we have.

\section{Sample selection and possible biases}
\label{sampleselection}
We cross-correlated the HST/ACS source catalog with the three redshift catalogs mentioned above to yield objects with spectroscopic redshifts in the range $0.2 \leq z \leq 1.2$. Since we have used spectroscopic redshifts from various surveys in the CDFS for the sample selection, our sample is liable to suffer from the same biases as those induced at the time of target selection by the spectroscopic surveys. The IMAGES survey is biased towards brighter galaxies with strong emission lines (Ravikumar et al. 2007). The shorter integration times used in the IMAGES survey results in poor SNR for fainter galaxies and results in difficulties in estimating redshifts for galaxies which exhibit only absorption lines or weak emission lines. Hence they are biased against faint, red early type galaxies. However, this bias is recovered to a large extent by the sample selection criterion of the GOODS/FORS2 redshift survey(Vanzella et al. 2005), which is tailor-made to exploit the red throughput and sensitivity of FORS2. This preferentially selects red, faint early types galaxies, and provides a nice complimentary redshift catalog to the IMAGES survey. Indeed this has been discussed at length by Ravikumar et al. \cite{ravi2007} (see their section 4.1) who have tested successfully the representativeness of the combination of IMAGES and FORS2 sample. The VVDS survey, on the other hand, is quite unbiased owing to the simple selection criterion of apparent magnitude $I_{AB} \leq 24.0$ and does not bias for or against any particular type of object (Le Fevre et al. 2004). The recovery success rate for the three redshifts surveys vary from 88\% for the VVDS, 77\% for GOODS/FORS2 and 76\% for the IMAGES survey. In addition to this, the above mentioned recovery rates are a function of magnitude of the object (see Le Fevre et al. 2004, Vanzella et al. 2005 \& Ravikumar et al. 2007). Therefore, as in any flux limited survey, fainter objects with weaker or no emission lines (eg. red absorption line galaxies) are underrepresented. This caveat must be kept in mind while interpreting the results presented in this paper.
Inspite of the limitations imposed by the flux-limited nature of the surveys, a combination of these three redshift catalogs yields a sample of galaxies which is reasonably representative of the field galaxy population out to a redshift of $\sim$1.2. 

\begin{figure}[h] \centering
   \includegraphics[angle=0, width=1.0\textwidth]{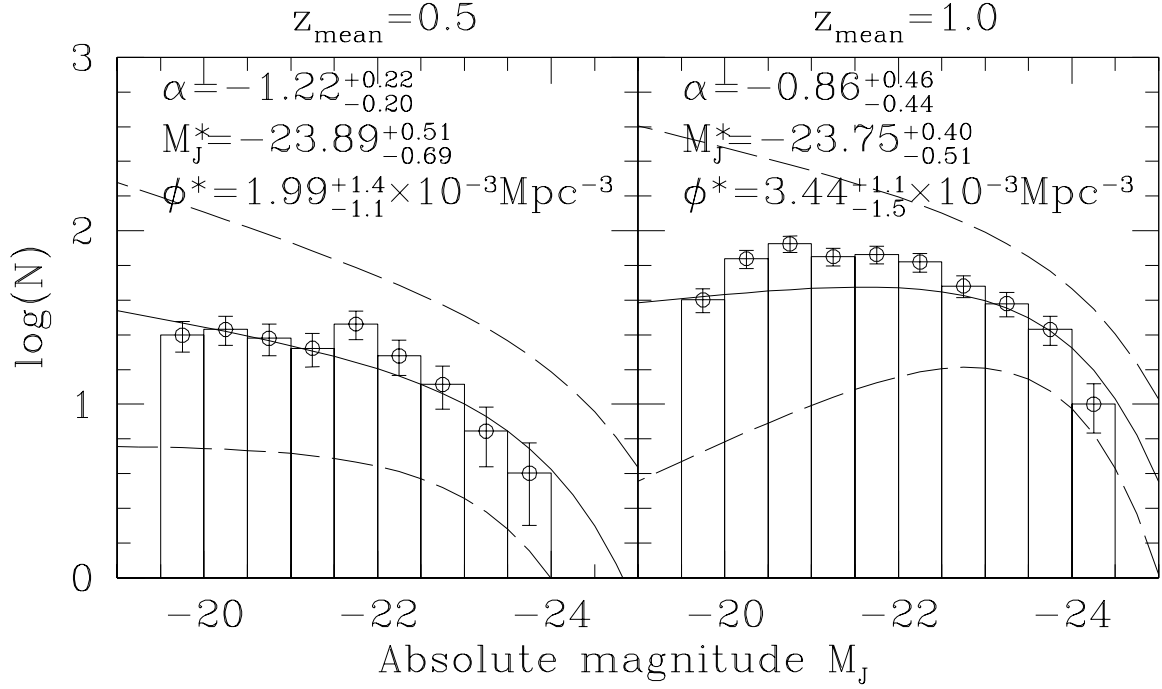}
   \caption{A comparison between the luminosity distribution of our sample galaxies in the redshift range $0.2 \leq z \leq 0.65$ (left panel) \& $0.65 \leq z \leq 1.2$ (right panel) respectively and the distribution predicted by the luminosity function of Pozzetti et al.~\cite{pozzetti2003}. The solid line is the luminosity distribution predicted by the mean values of $\alpha$, $M^{*}$ and $\phi^{*}$ given by Pozzetti et al.~\cite{pozzetti2003}, whereas the two dashed lines on either side demarcate the 1 sigma region around the mean as parametrized by the errorbars on $\alpha$, $M^{*}$ and $\phi^{*}$ mentioned above.}
   \label{lumfn}
\end{figure}

Galaxies with rest frame absolute magnitude brighter than $M_J(AB)=-19.5$ have been selected as the primary galaxies, leaving us with 695 galaxies. The luminosity cut off was selected to ensure that the sample does not suffer from any incompleteness (Malmquist bias) in the highest redshift bin. Rest-frame J band is significantly less affected by sporadic star formation than bluer bands, and is more representative of the stellar mass content. Fig.~\ref{lumfn} illustrates that the resulting sample is rather well representative of the Schechter luminosity function from Pozzetti et al.~\cite{pozzetti2003} at mean redshift $z_{mean}\sim0.5$ \& $1.0$. The Schechter luminosity function is parametrized as 
\begin{eqnarray}
 \phi(M) =   0.9210~\phi^{*}10^{0.4(\alpha+1)(M^{*} - M)}\times e^{-10^{0.4(M^{*} - M)}}
\label{schechter}
\end{eqnarray}

where the values of $\alpha$, $M^{*}$ and $\phi^{*}$ from Pozzetti et al.~\cite{pozzetti2003} are given in Fig.~\ref{lumfn}.

%This ensures that our sample galaxies are representative of the general field galaxy population with $M_J\leq-19.5$ and do not suffer from incompleteness in any luminosity range. 

\begin{figure}[h] \centering
   \includegraphics[angle=0, width=1.0\textwidth]{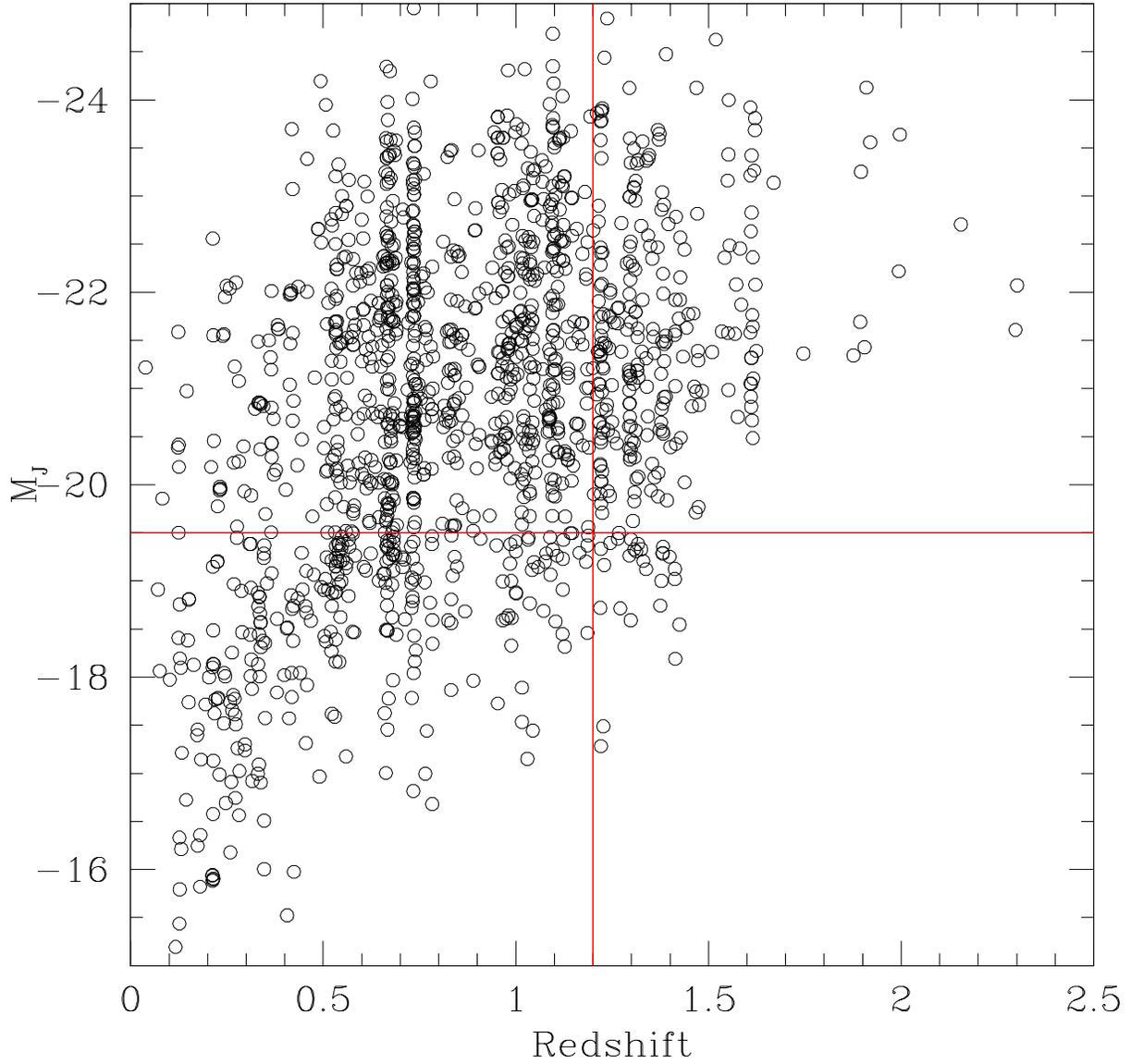}
   \caption{A pictorial representation of the sample selection criterion applied by us in terms of redshift and luminosity cutoffs. Notice the prominent large scale structure at $z\sim0.735$}
   \label{mj_completeness}
\end{figure}

Fig.~\ref{mj_completeness} shows pictorially the sample selection criterion applied by us in terms of redshift and luminosity cutoffs.

\section{$K_{S}$ band photometry}
\label{ks}

Near-infrared imaging observations of the CDFS have been carried out, as part of GOODS, in J, H, \& Ks bands using VLT/ISAAC. In the present work, we have used the ESO GOODS/EIS Release Version 1.5 (Vandame et al., in preparation). This data release includes 24 fully reduced VLT/ISAAC fields in J \& Ks bands, covering 159.1 and 159.7 $\arcmsq$ of the GOODS/CDFS region, respectively.  

\begin{figure}[h] \centering
   \includegraphics[angle=0, width=1.0\textwidth]{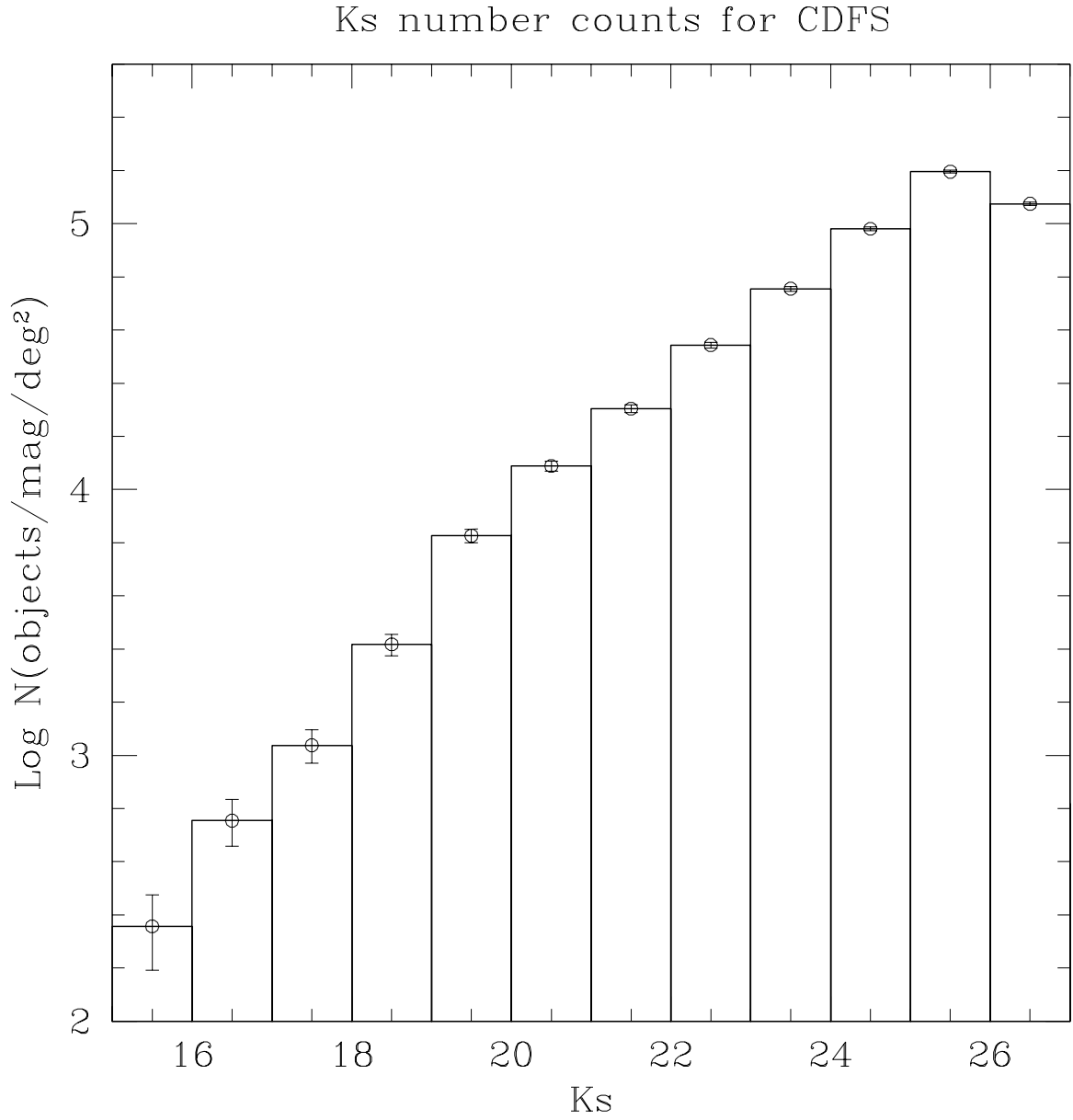}
   \caption{The Ks band differential number counts obtained using the SExtractor catalog that we have constructed.}
   \label{numbercounts}
\end{figure}

We have carried out photometry with SExtractor on the J \& Ks images, using DETECT\_THRESH=1.0 ; DETECT\_MINAREA=5 and DEBLEND\_MINCONT=0.0005 in both the filters. Deblending is a major source of worry when studying close pairs of objects. However, we are essentially immune to this problem as our selection criteria demands that the fainter member of a pair have at least 25\% of the flux of the other member to qualify as a major pair. Hence even if SExtractor ends up splitting a source into two (or more) owing to a low value of DEBLEND\_MINCONT, it will still not get counted as a legitimate major pair. The combination of DETECT\_THRESH, DETECT\_MINAREA and DEBLEND\_MINCONT has been optimised after several iterations of experimenting with different values for these parameters and subsequent visual examination after overlaying the derived catalog onto the J \& Ks band images. The differential number counts obtained using our SExtractor catalog are given in Fig.~\ref{numbercounts}.
% and were found to be in good agreement with the number counts published by Djorgovski et al.~\cite{djorgovski1995}.

\section {Finding pairs of galaxies}
%\labsecn{pairs}
\label{pairs}

We have used a simple prescription for finding pairs of galaxies by identifying neighbours
%using the methodology defined by earlier workers in the field (Le Fevre et al.~\cite{lefevre2000}, Bundy et al.~\cite{bundy2004}). We look for galaxies 
within a projected radius of $20h_{100}^{-1}$\kpc~ for each of the 695 main galaxies. For each neighbour, we find the difference in the magnitude $\delta m$ of the main galaxy and the neighbour in both $z$ as well as $K_s$ filter separately. If the condition $-1.5 \leq \delta m \leq 1.5$ is satisfied in at least one of the two filters, the pair is designated to be a tentative {\it{major pair}} in that particular filter. This condition ensures that the fainter member of the pair has at least 25\% of the $z$ or $K_s$ band flux of the brighter member. It is important to note here that it is entirely possible for a neighbour to be {\it{brighter}} than the primary galaxy. It may indeed happen that a fainter galaxy within a pair has a spectroscopic measurement, while the brighter galaxy has none, owing to observing constraints of the redshift survey. 
 %It is one of the vagaries of redshift surveys that sometimes a fainter object has it's spectrum taken, but brighter sources in the neighbourhood might not be observed spectroscopically owing to observing constraints. 
%Hence it is important while looking for major merger candidates to search over the whole range $-1.5 \leq \delta m \leq 1.5$. 
%and not just look for neighbours which are fainter than the host galaxy. 
Not including brighter neighbours can lead to an underestimation of the derived merger fraction. 
We find that not including major pairs in which the neighbour is brighter than the primary galaxy, we will underestimate the number of major pairs by 34\% in the Ks filter and 24\% in the z filter. The importance of including brighter companions in the definition of major pairs has been missed by earlier workers in the field. 

\begin{figure*}[t]
 \includegraphics[height=5.8cm,clip]{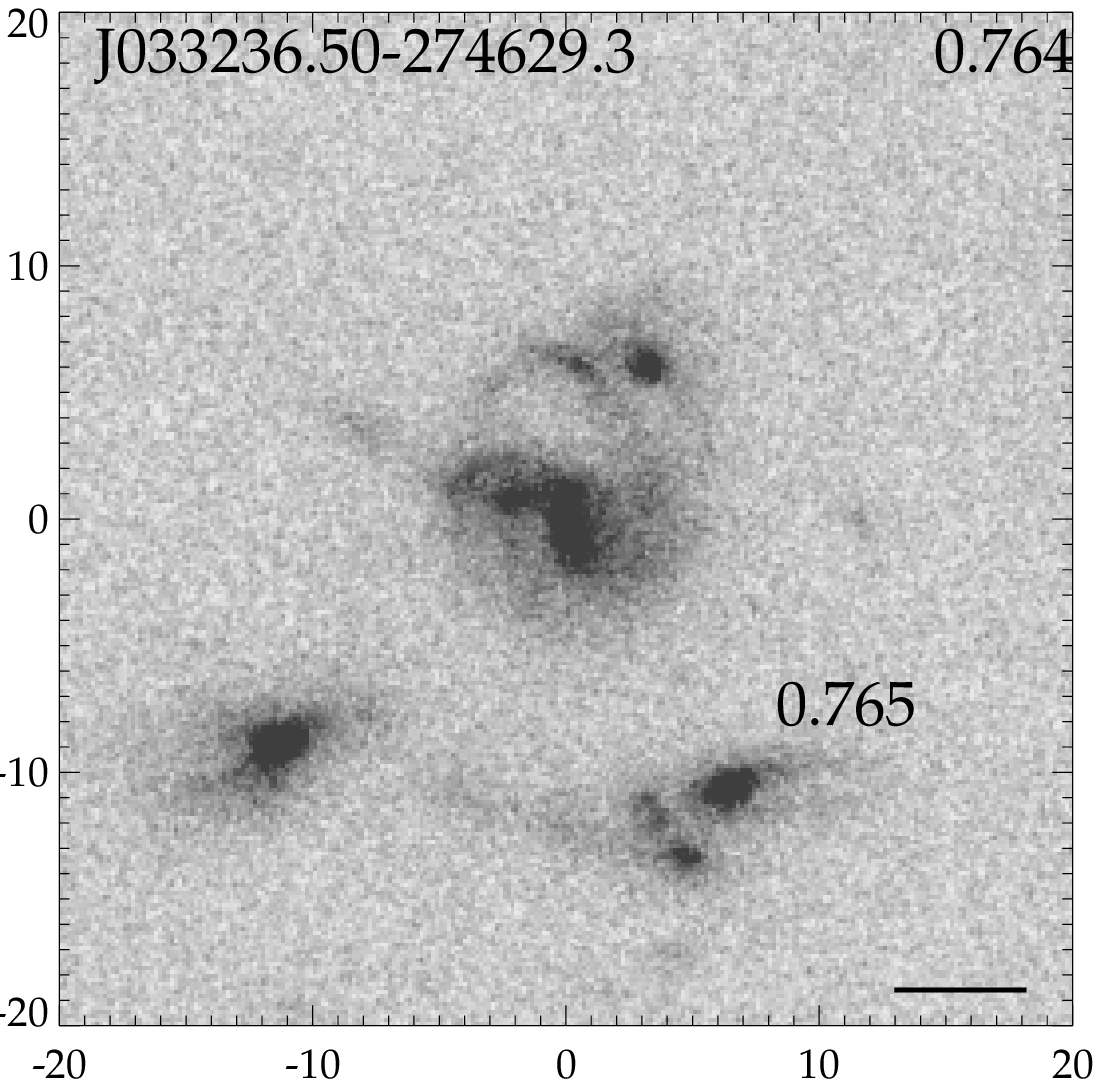}%
 \includegraphics[height=5.8cm,clip]{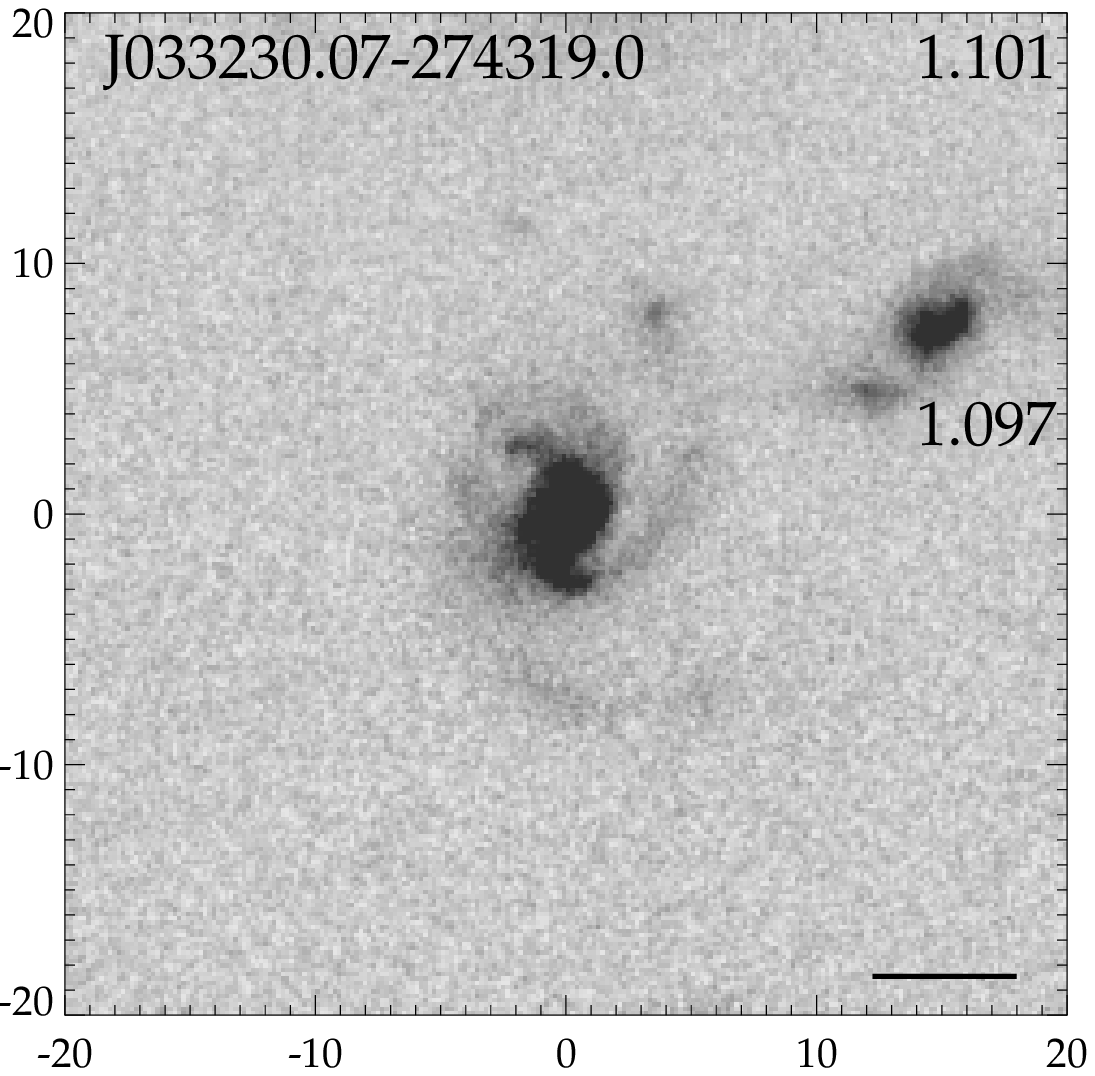}%
\includegraphics[height=5.8cm,clip]{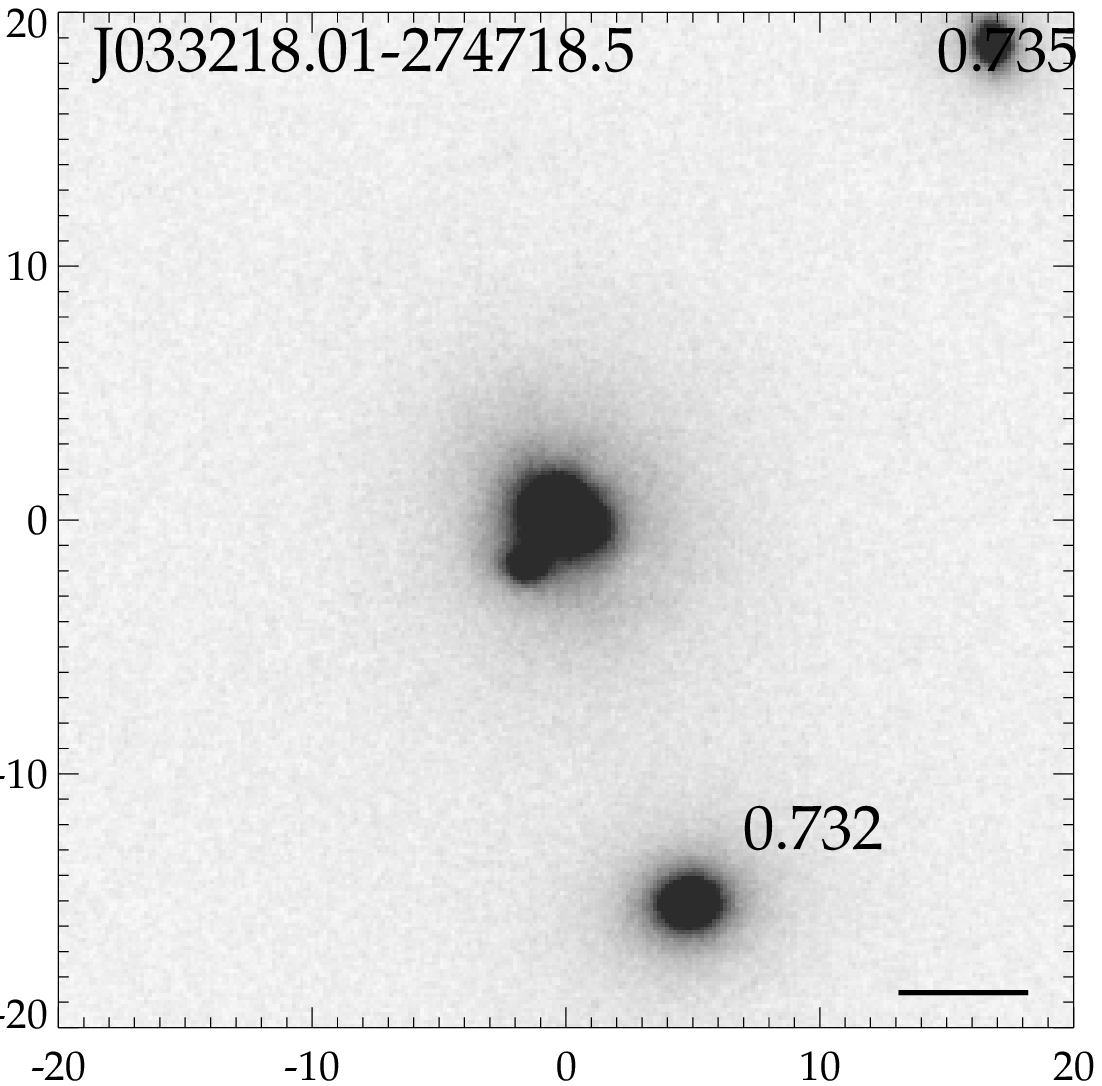}
\includegraphics[height=5.8cm,clip]{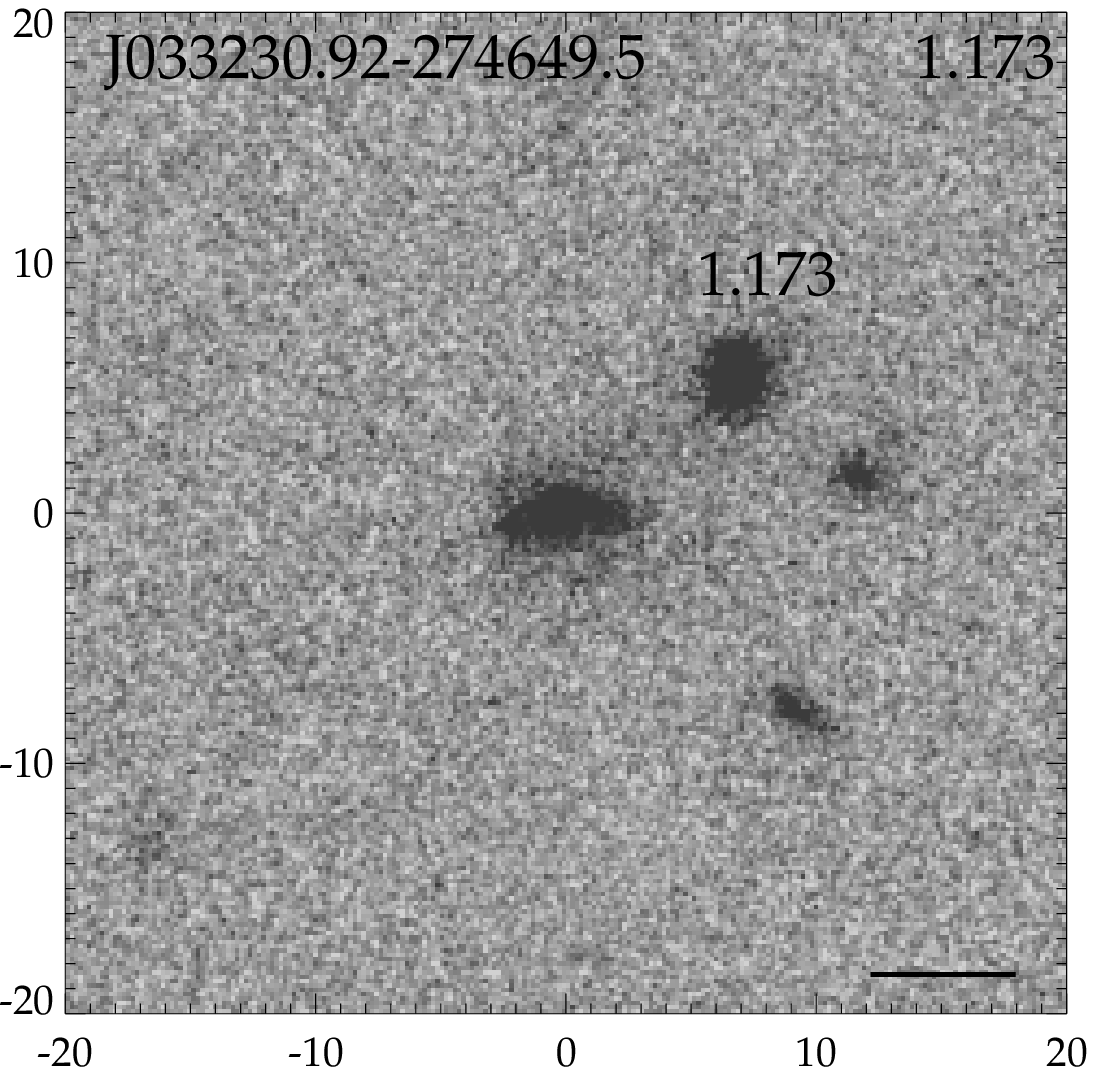}%
\includegraphics[height=5.8cm,clip]{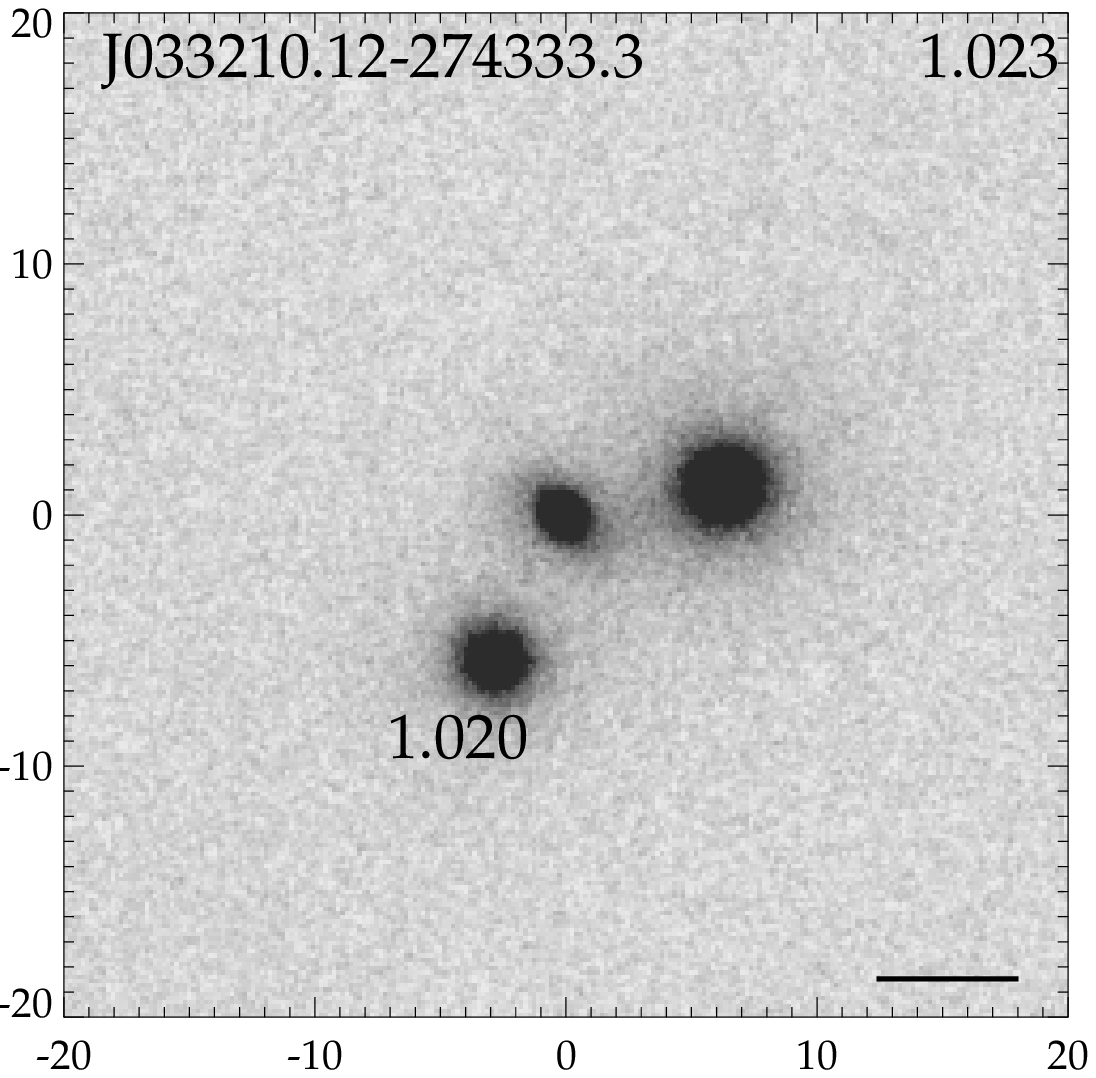}%
 \includegraphics[height=5.8cm,clip]{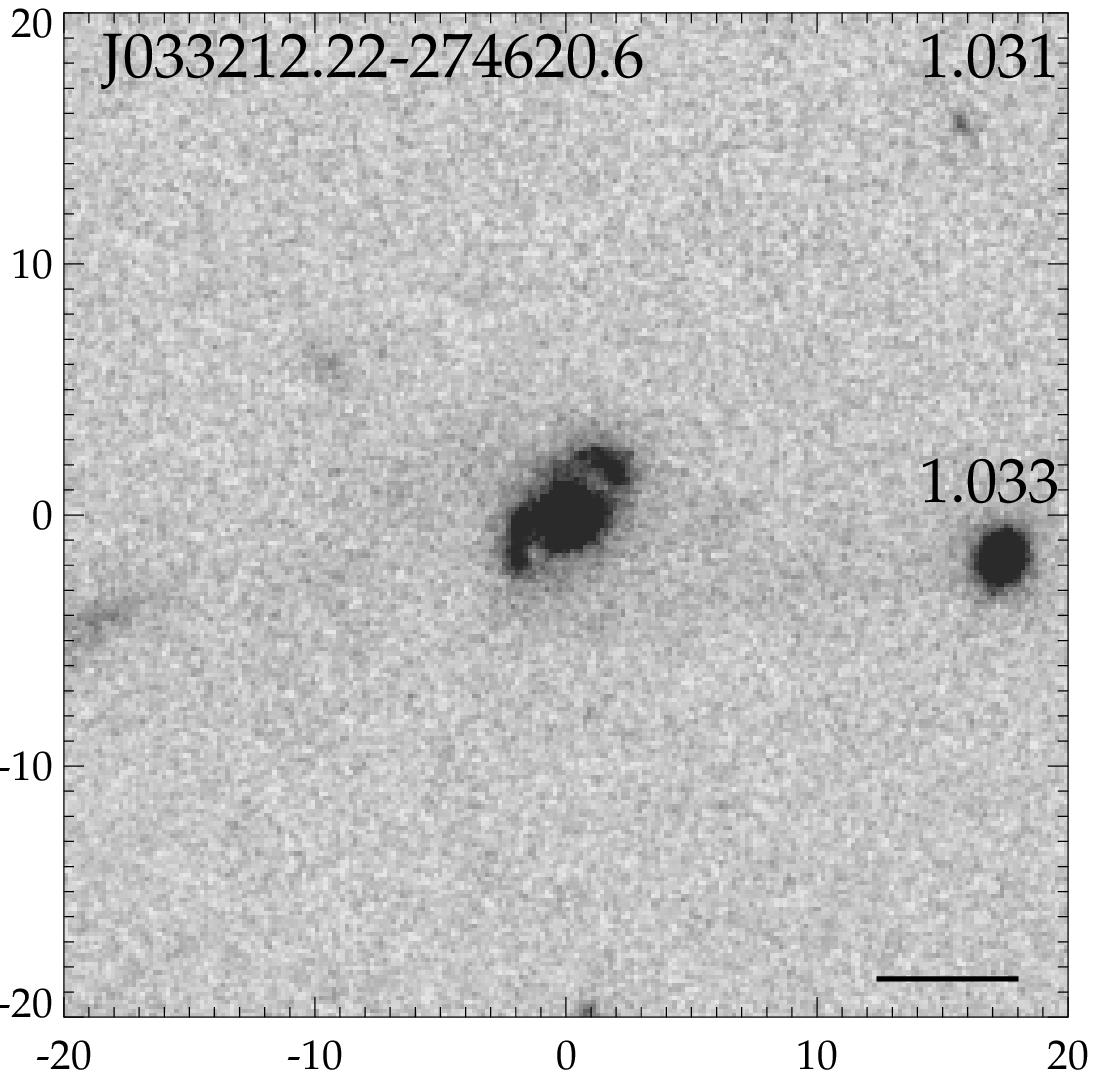}
 \includegraphics[height=5.8cm,clip]{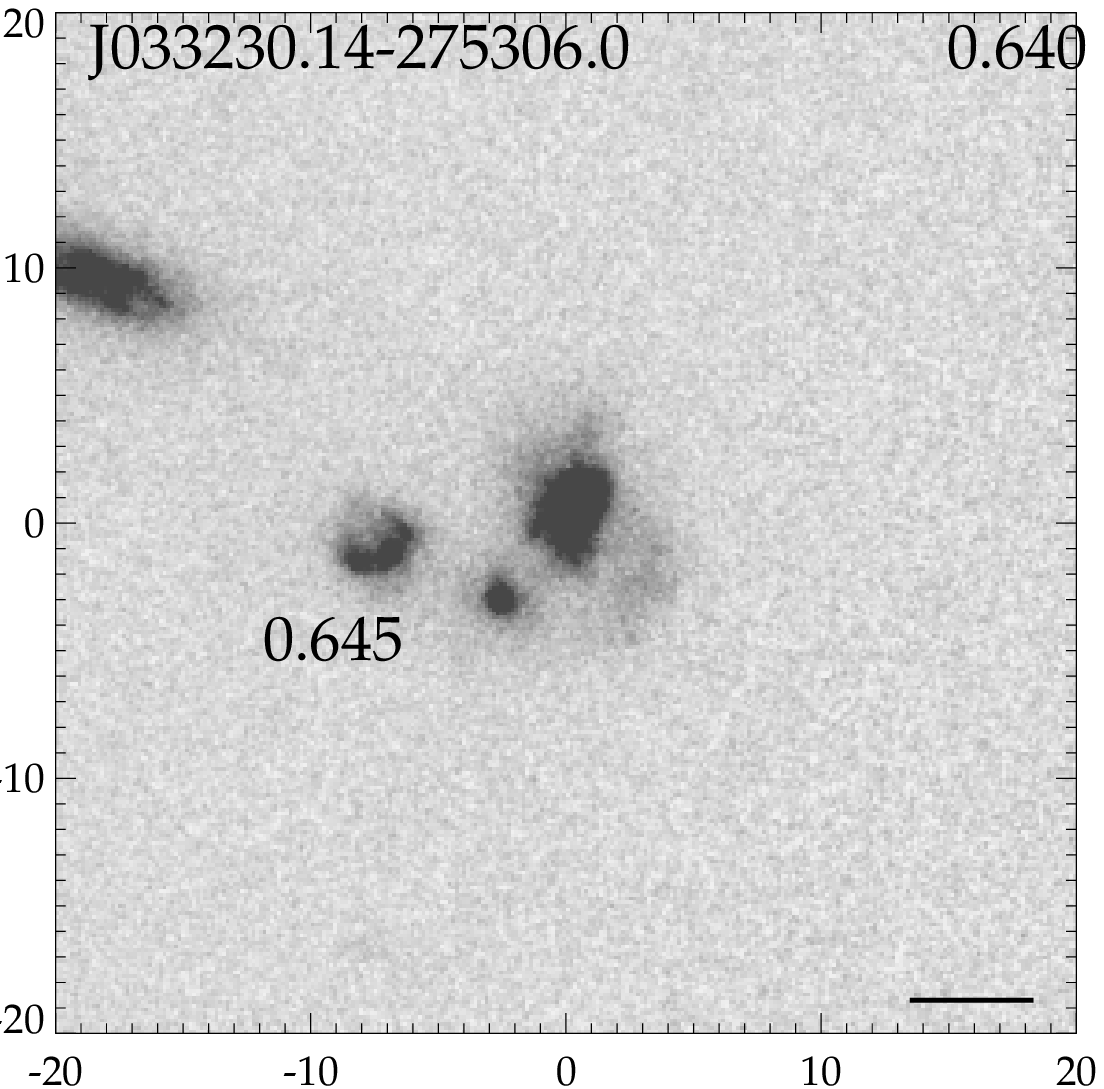}%
\includegraphics[height=5.8cm,clip]{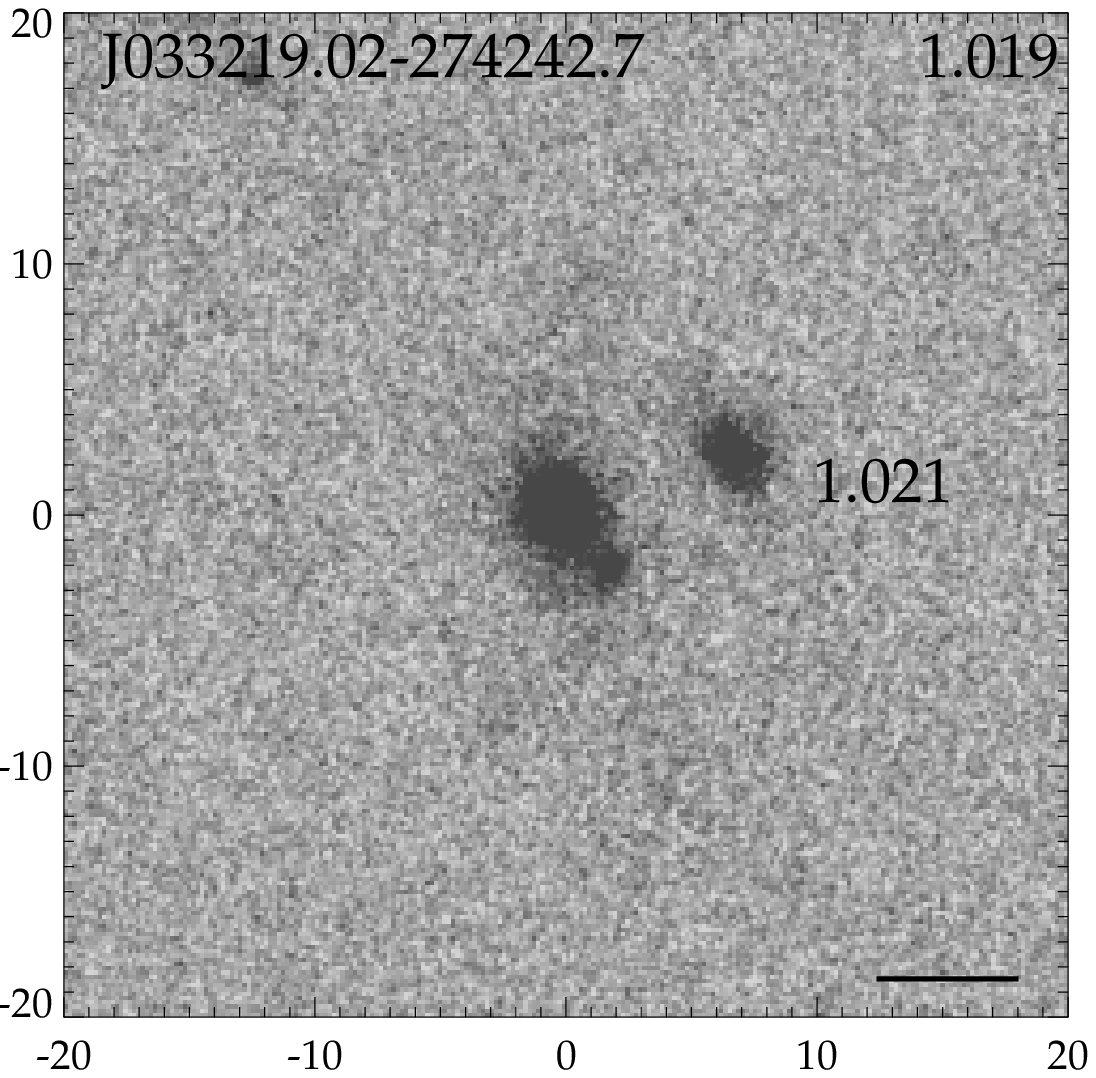}
\caption{A montage of $z$ band HST/ACS images of some of the confirmed major pairs, identified in the $z$ filter, where the secondary member has the same spectroscopic redshift as the host galaxy. Each image is $40 \times 40 h_{100}^{-1}$ \kpc~ in size. The unique object ID of the host galaxy is marked on the top left of each image, the redshift of the host galaxy is marked on the top right and the redshift of the neighbour is marked next to it. The horizontal bar on the bottom right shows a scale of 1 arcsec.}
\label{montage}
\end{figure*}

\begin{figure*}[t]
\includegraphics[height=5.8cm,clip]{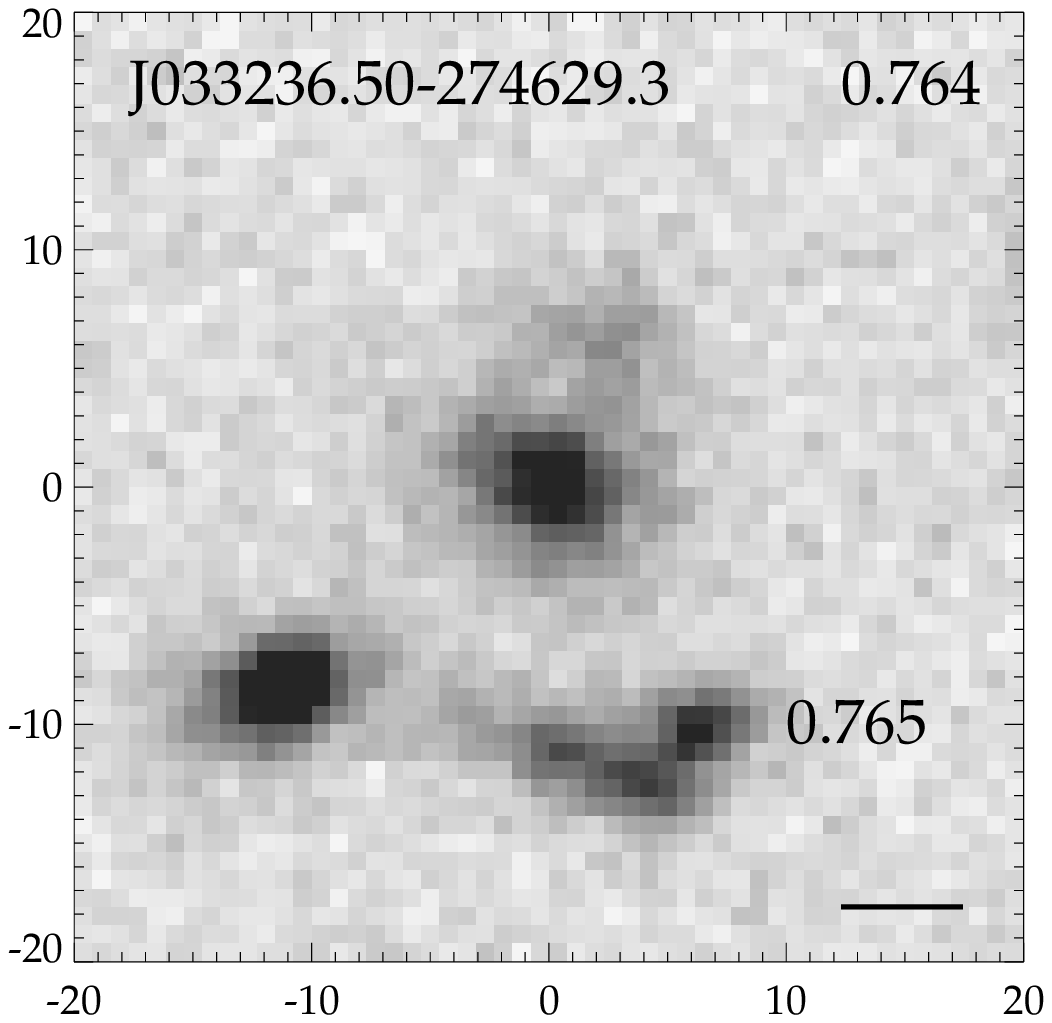}%
 \includegraphics[height=5.8cm,clip]{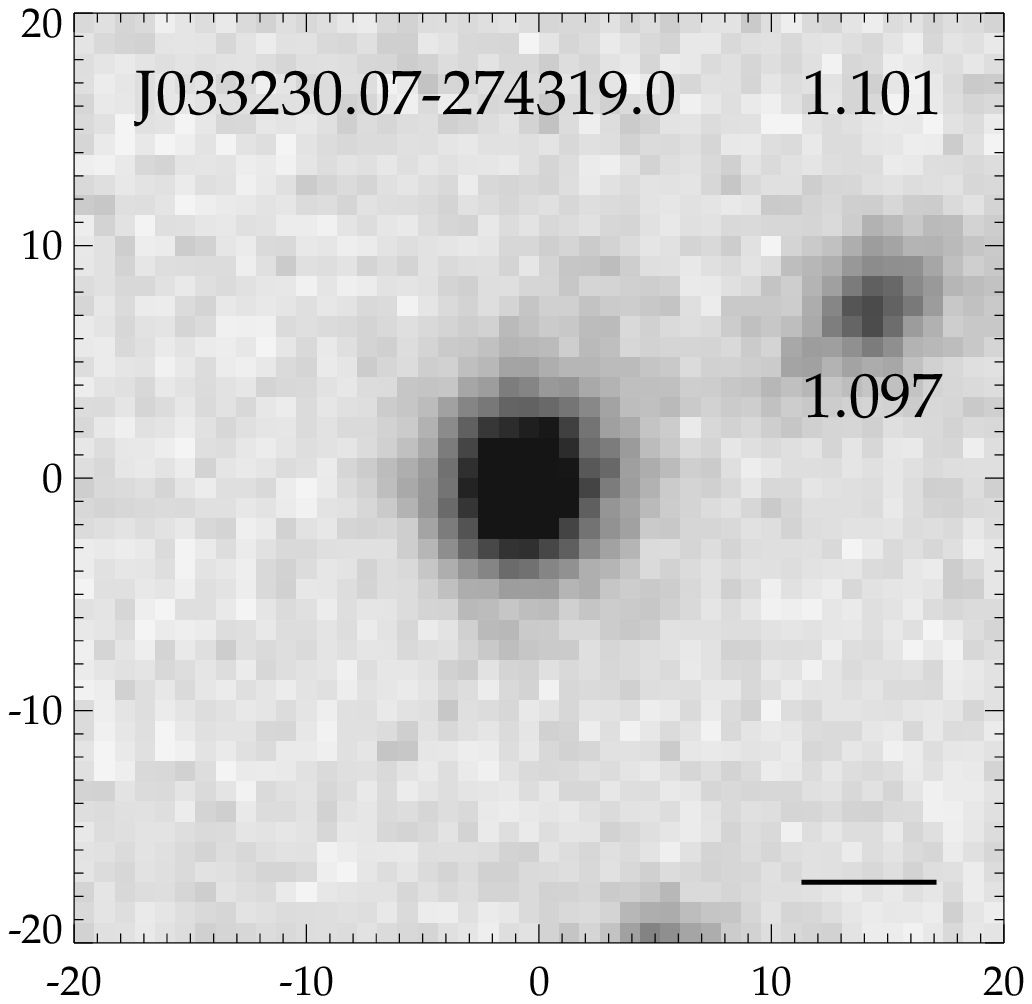}%
 \includegraphics[height=5.8cm,clip]{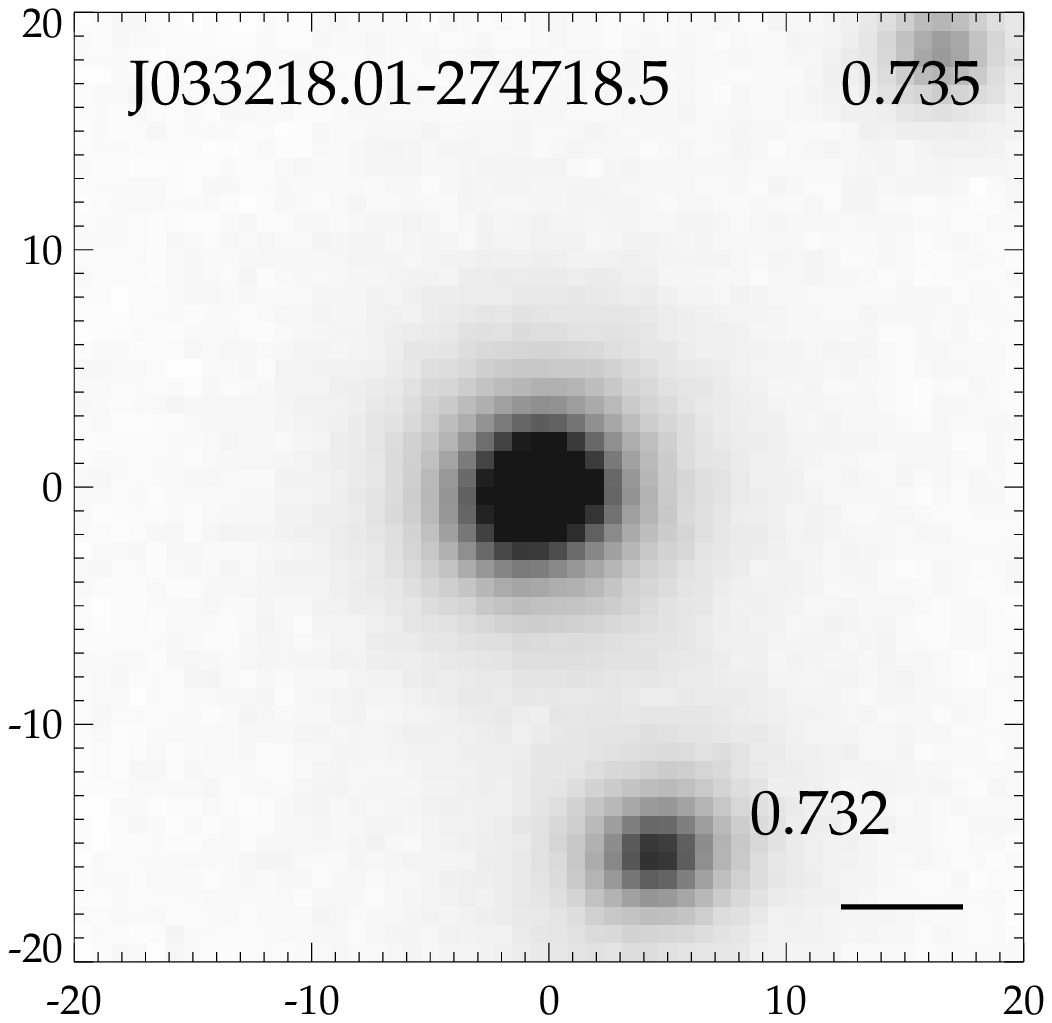}
 \includegraphics[height=5.8cm,clip]{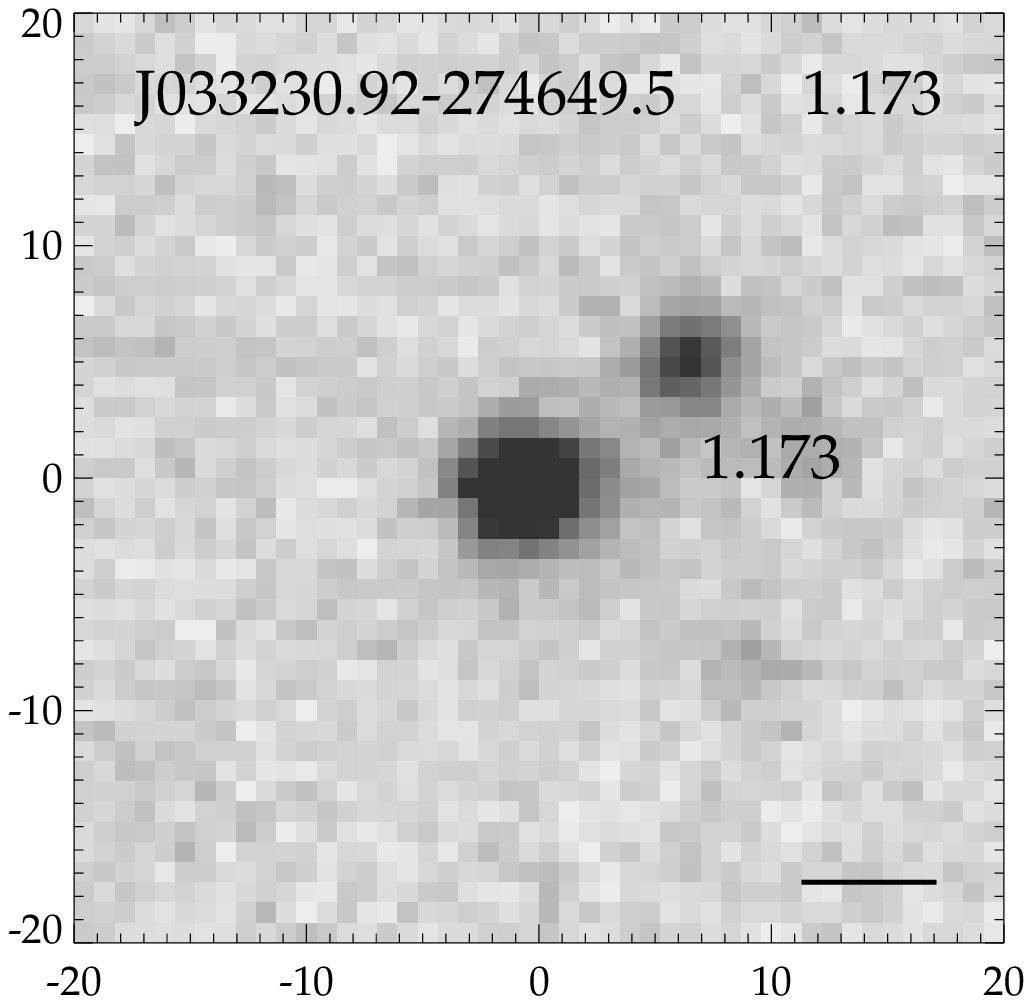}%
\includegraphics[height=5.8cm,clip]{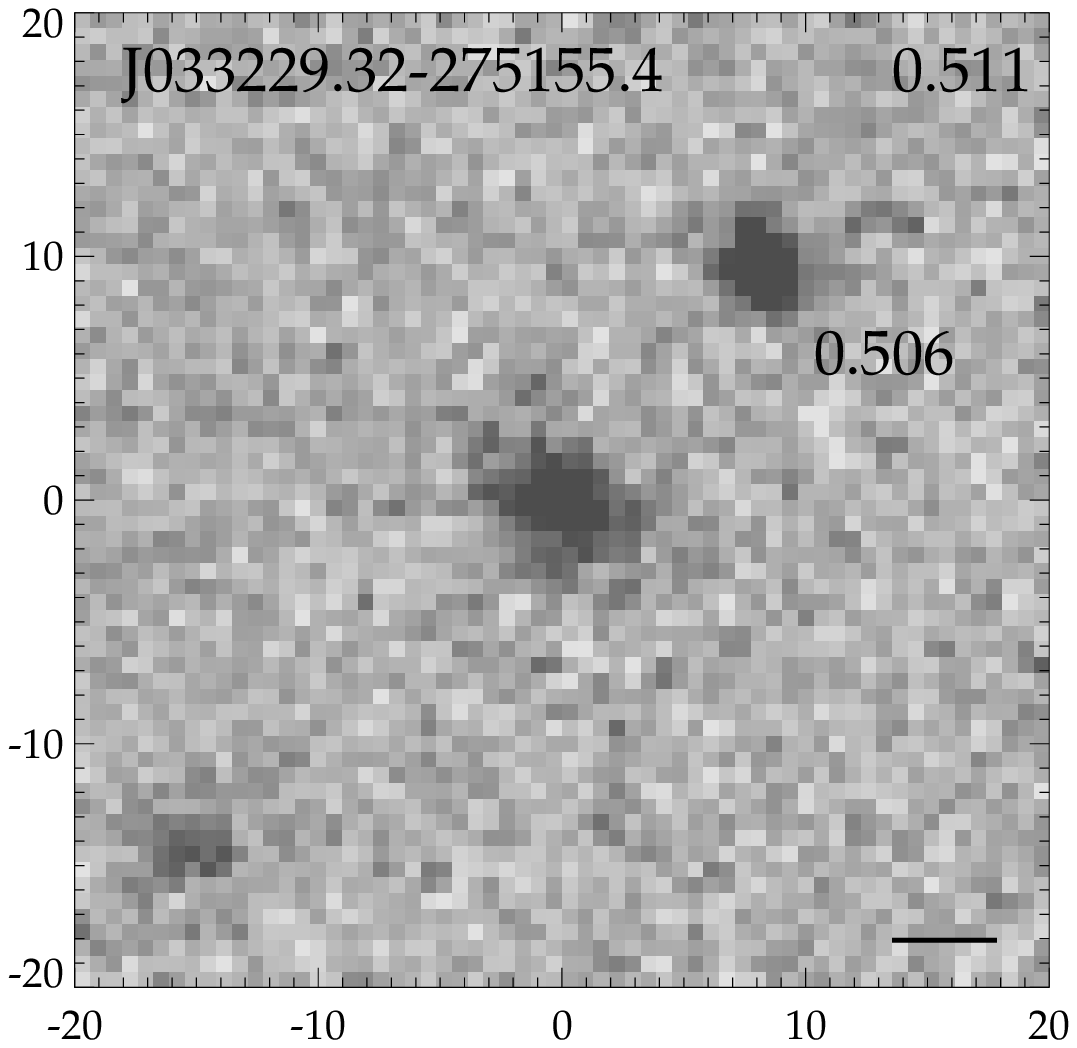}%

\caption{A montage of $K_s$ band images of some of the confirmed major pairs, identified in the $K_s$ filter, where the secondary member has the same spectroscopic redshift as the host galaxy. Each image is $40 \times 40 h_{100}^{-1}$ \kpc~ in size. The unique object ID of the host galaxy is marked on the top left of each image, the redshift of the host galaxy is marked on the top right and the redshift of the neighbour is marked next to it. The horizontal bar on the bottom right shows a scale of 1 arcsec.}
\label{montage_k}
\end{figure*}

The simple selection criterion employed by us yielded a sample of 162 and 226 tentative major pair candidates in the $K_s$ filter, and in the $z$ filter, respectively. We do not have the redshift for the secondary member of a pair in most cases, therefore some of them can simply be foreground/background superpositions. Such a contamination can be statistically corrected for by using number counts of objects in the field and is explained in section~\ref{contamination}. However, in the cases where the secondary member does have a redshift, we have used it to rule out those apparent pairs from the list which are obviously at different redshift as compared to the host galaxy. This left us with 125 and 181 major pair candidates in the $K_s$ filter, and in the $z$ filter, respectively. Furthermore, a lower limit of $5h_{100}^{-1}$\kpc~ is imposed on the separation between the major pair candidates and the main galaxy in order to avoid confusing clumpy extensions of the main galaxy itself as a neighbour. This left us with 119 and 170 major pair candidates in the $K_s$ filter, and in the $z$ filter, respectively.
These major pair candidates, are then segregated into three redshift bins and are listed in Table~\ref{table1}. Fig.~\ref{montage} shows a montage of $z$ band HST/ACS images of some of the confirmed major pairs, identified in the $z$ filter, where the secondary member has the same spectroscopic redshift as the host galaxy. Similarly, Fig.~\ref{montage_k} shows $K_s$ band images of some of the confirmed major pairs identified in the $K_s$ filter. Note that there is a one-to-one correspondence between the first four objects in Fig.~\ref{montage} and Fig.~\ref{montage_k}.

We have used the $z$ as well as the $K_s$ filter to define major mergers because the former filter allows to find companions very near to the primary galaxy (thanks to the exquisite spatial resolution of the ACS camera) and the latter is less affected by star formation and more representative of the stellar mass (e.g. Bundy et al. 2004).
% the underlying stellar population of a galaxy (and therefore of it's mass) than optical flux and is therefore less likely to be biased by a recent burst of star formation that might have been triggered by the putative interaction as pointed out by Bundy et al.~\cite{bundy2004}. 
 
%

%

%\begin{figure}[h!]
%\includegraphics[height=4.3cm,clip]{923844_1_compress.ps}%
% \includegraphics[height=4.3cm,clip]{240864_Ks_1_compress.ps}
%\caption{An example of a pair of galaxies at redshift 0.273 in the HST/ACS $z$ filter (left) and in $K_s$ filter (right). Both images are 18 arcsec in size. The circle in each figure has a radius of $20 h_{100}^{-1}$\kpc. Notice the faint tidal features visible in the images. The strong dust lane visible in the F850LP filter in one of the galaxies, disappears in the $K_s$ image.}
%\label{pair}
%\end{figure}

%\begin{figure}[h!]
%\includegraphics[height=4.5cm,clip]{lum_fn1_tsr.ps}%
%\includegraphics[height=4.5cm,clip]{lum_fn2_tsr.ps}
%\caption{}
%\label{lumfn}
%\end{figure}

\subsection{Estimation of the foreground/background contamination}
\label{contamination}

Since we do not have redshift for the secondary member of a pair in most cases, some of them can simply be foreground/background superpositions. Such a contamination can be statistically corrected for by using number counts of objects in the field in the z or $K_s$ filter to estimate the probability of finding an object within projected $5h_{100}^{-1}\kpc\leq r \leq 20h_{100}^{-1}$\kpc~ of the main galaxy and within $\pm1.5$ magnitude of the main galaxy. This statistically expected number can then be subtracted off the number of neighbours observed around a given host galaxy to yield the number of true neigbours for that particular host. 
Mathematically, this statistical correction is calculated as

\begin{eqnarray}
    \int_{m-1.5}^{m+1.5}N(m')dm' \times \pi (r_{20h_{100}^{-1}\kpc}^{2} - r_{5h_{100}^{-1}\kpc}^{2} )          \label{integrate}
\end{eqnarray}

where, N(m') is the differential number counts of objects in a given filter, m is the apparent magnitude of the host galaxy in the same filter, and $r_{20h_{100}^{-1}\kpc}$ and $r_{5h_{100}^{-1}\kpc}$ are the projected radius in degrees corresponding to $20h_{100}^{-1}\kpc$ and $5h_{100}^{-1}\kpc$ at the redshift of the main galaxy.
In the $K_s$ band, these number counts were established by using the SExtractor catalog as explained in section~\ref{ks}. In the $z$ band, we have estimated the number counts using the catalog published by the GOODS team.    

Equ.~\ref{integrate} gives us the number of sources that are expected to be present within $\pm1.5$ magnitude and within projected $5h_{100}^{-1}\kpc\leq r \leq 20h_{100}^{-1}$\kpc~ of the host galaxy by pure chance coincidence.
It may be noted that this statistical correction is only calculated (and subtracted) in those
cases where the neighbor does not have a spectroscopic redshift. 
If the neighbor has a spectroscopic redshift, we know for sure if it is a real pair or not, hence no need
to apply the statistical correction.

%It is only in cases where the neighbor does not have a specz, that there exists an uncertainty as to whether the object is really associated with the host or is a chance superposition. Hence it is only in these cases that we need to subtract off the statistical correction that we have calculated using number counts. 
This number is calculated for every host galaxy in our sample which has a neighbour without a redshift. The final number is listed in Table~\ref{table1}, segregated into three redshift bins, summed over all the host galaxies (with neighbours sans redshift) in a given redshift bin.

\subsection{Test of our contamination calculation}
\label{test}

We tested the accuracy of our statistically estimated foreground/background contamination derived in Section~\ref{contamination} in two ways. First, we compute the ratio of the statistically expected neighbours to the number of major neighbours actually found in projection as listed in Table~\ref{table1}. This can be termed as a sort of {\it{contamination rate}}. This contamination rate is then compared to earlier works of Le Fevre et al.~\cite{lefevre2000}. The contamination rate that we find here (in the z band) is $42\% \pm 6.3\%$. This is in excellent agreement with the contamination rate of $49\% \pm 13\%$ (as derived from Table 3 of Le Fevre et al. 2000). This agreement gives us additional confidence in our estimate of the foreground/background contamination.

\begin{figure}[h] \centering
   \includegraphics[angle=0, width=1.0\textwidth]{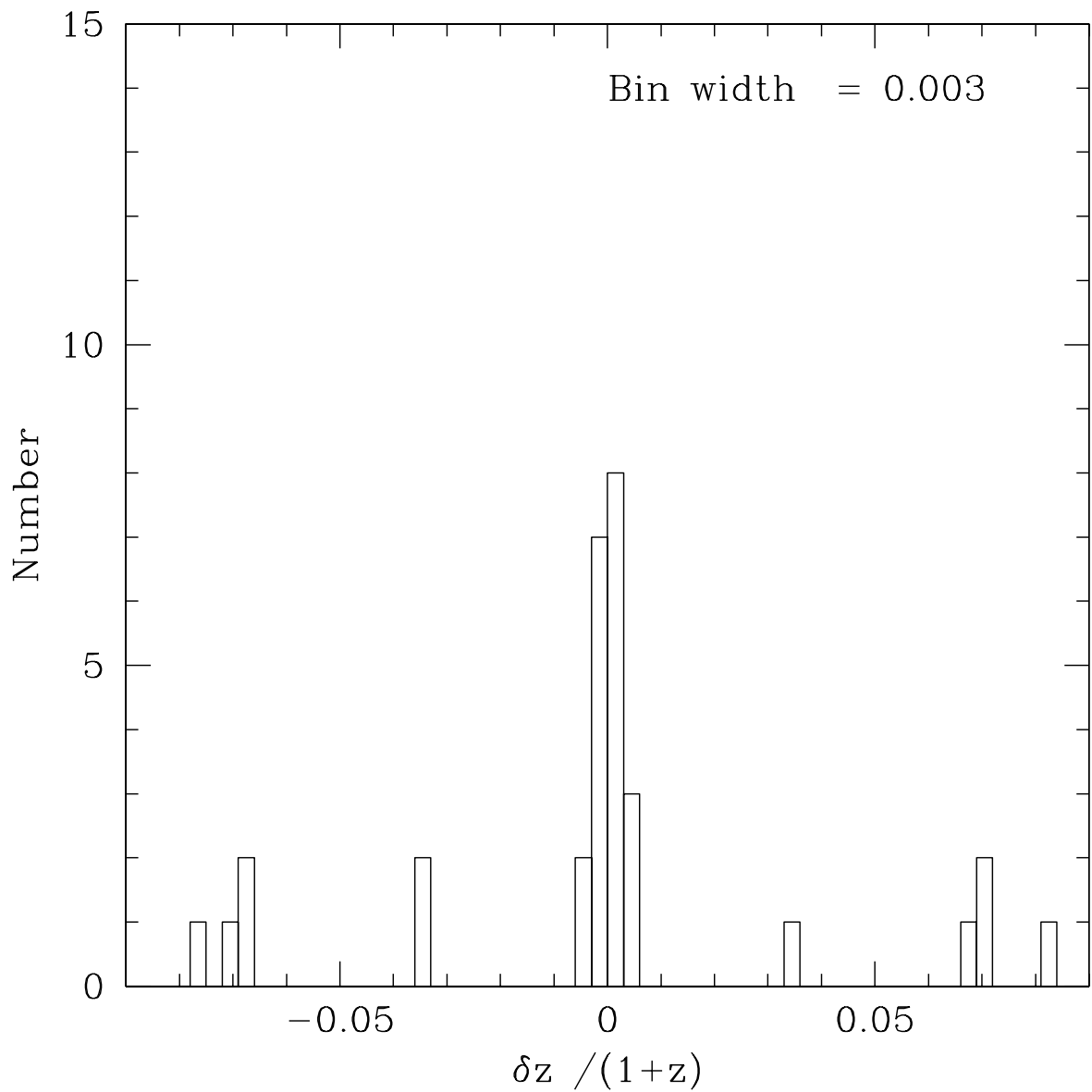}
   \caption{Histogram of $(z_{host}-z_{neigh})/(1+z_{host})$ for the 65 major pairs where both members have a spectroscopic redshift. Note that only 20 pairs have $-0.006 \leq \delta z/(1+z) \leq 0.006$ as seen in the figure. Most other pairs have $\delta z/(1+z)$ much larger than 0.006 so that they are out of the frame. }
   \label{deltaz}
\end{figure}

In addition to comparison with earlier works, we also checked our contamination rate of $42\% \pm 6.3\%$ with the contamination rate derived using the subset of our major pairs where we have spectroscopic redshifts for both members of the pair. There are 65 major pairs in the z band where both members of the pair have spectroscopic redshifts. In order to establish how many of these 65 major pairs are indeed at the {\it{same}} redshift, we plotted a histogram of $\delta z/(1+z) = (z_{host}-z_{neigh})/(1+z_{host})$ for these 65 major pairs. This is shown in Fig.~\ref{deltaz}. It turns out that there are a number of pairs with $-0.006 \leq \delta z/(1+z) \leq 0.006$, which are presumably located at the {\it{same}} redshift, with the residual difference in redshift $\delta z/(1+z)$ being attributed to redshift measurement errors and peculier velocities of the members. 

A $\delta z/(1+z)$ = 0.006 corresponds to delta velocity of 1800km/sec. We were forced to use this seemingly large value of $\delta z/(1+z)$ due to the errorbars on the spectroscopic redshifts of the objects. The quoted errorbars in the redshift measurements are of the order $\sim$0.002 for VVDS and GOODS/FORS2 and $\sim$0.006 for the IMAGES survey. In addition to this, these quoted errorbars are to be treated as lower limits due to the way in which they are calculated (see eg. Vanzella et al. 2005). Since we are using three different redshift surveys, we have taken to using the worst errorbars i.e. $\delta z/(1+z)$ = 0.006. This implies that the delta velocity of 1800 km/sec is to be treated as an upper limit, and the true delta velocity is liable to be smaller.

In Fig.~\ref{deltaz} many objects have $\delta z/(1+z)$ much larger than 0.006 (so much so that most of them are out of the frame in Fig.~\ref{deltaz}). The members of such pairs are treated as having {\it{discordant redshifts}} and are essentially foreground/background superpositions. From this plot, we can then derive the fraction of pairs that are actually foreground/background superpositions (i.e. the {\it{contamination rate}}). This {\it{contamination rate}} is found to be $\sim 69\% \pm 13\%$.
On first sight this might appear larger than the {\it{contamination rate}} of $42\% \pm 6.3\%$ that we have obtained using our statistically estimated foreground/background contamination (although notice the large errorbars since we are dealing with small number statistics here). However, we checked the spectra of the pairs having discordant redshifts ourselves. It turns out that in many cases, the redshift determination can be quite poor indeed, specially in the low exposure and low resolution VVDS (as parametrised by their redshift flag). So in cases where the neighbor has a VVDS redshift flag 2 (75\% confidence) or worse, there is a much greater risk of the pair being designated as having discordant redshifts. In addition to this, one also has to allow for the fact that some of the pairs using redshifts from the other two redshift catalogs can also be erroneously counted as discordant (for the same reason as above). Such pairs account for $\sim$15/45 discordant redshift pairs in the z band. This can significantly swing the number of pairs with discordant redshifts. If we discard these 15 discordant cases and recalculate the contamination rate, we get $60\% \pm 14\%$ as the contamination rate, which is in much better agreement with the {\it{contamination rate}} of $42\% \pm 6.3\%$ that we have obtained using our statistically estimated foreground/background contamination. 

In the nutshell, our contamination calculation is robust and the contamination rate compares well with the contamination rate reported in literature, as well as with the subsample of our major pairs where we have spectroscopic redshifts for both members of the pair. Also, as a result of our eagerness to throw out prospective pairs as having discordant redshifts due to spurious redshift determination in some cases, we are going to end up with an estimate of merger fraction which is a bit too low and thus it should be treated as a lower limit.

\subsection{Source blending and photometric completeness}
%\labsubsecn{blending}
\label{blending}

\begin{figure*}[t] \centering
  \includegraphics[angle=0, width=1.0\textwidth]{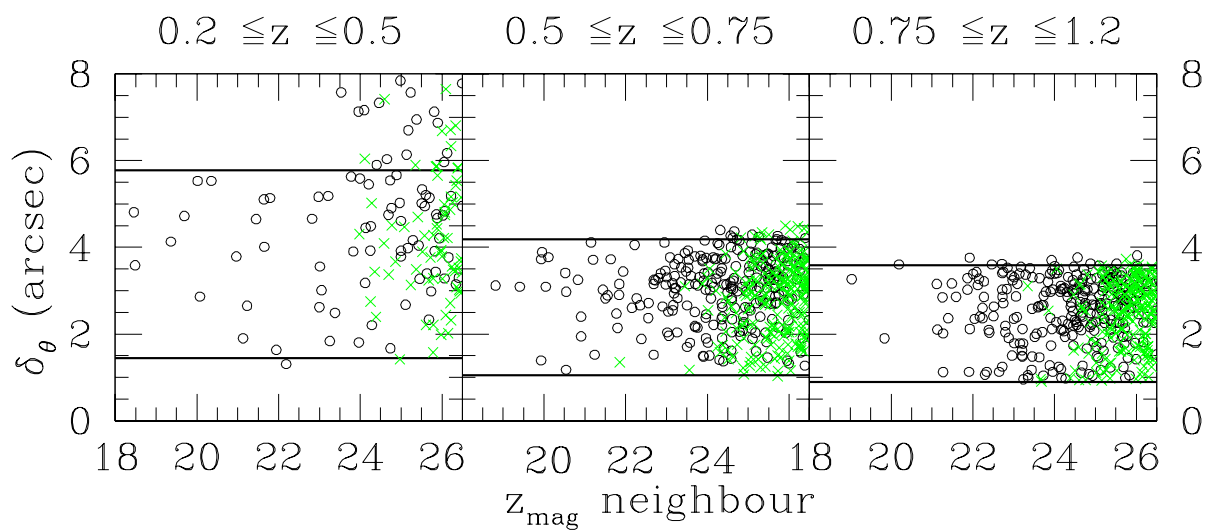}
  \caption{The angular separation between the main galaxy and the neighbour identified in the $z$ filter Vs the $z$ magnitude of the neighbour, segregated into three redshift bins as indicated. The open circles are neighbours which have measured $K_s$ magnitude, whereas the green crosses are neighbours which do not have $K_s$ magnitude. The horizontal lines correspond to $20h_{100}^{-1}\kpc$ and $5h_{100}^{-1}\kpc$ at the mean redshift of each bin.}
   \label{theta}
\end{figure*}

A major bias when studying close pairs of objects may be caused by the fact that sources are blended when their separation is comparable to the PSF. It particularly affects ground based $K_s$ images because the $K_s$ PSF ($\sim$0.5\arcsec) is much larger than the HST optical PSF ($\sim$0.1\arcsec). This is illustrated in Fig.~\ref{theta}, where we have plotted the angular separation between the main galaxy and neighbours identified in the $z$ filter against the $z$ magnitude of the neighbour, segregated into three redshift bins. 
A neighbour here is identified as any galaxy within $5h_{100}^{-1}\kpc\leq r \leq 20h_{100}^{-1}$\kpc~ of the primary galaxy, regardless of it's magnitude. The resultant list of 1085 neighbours includes major as well as minor neighbours, and constitutes the pool from which the major pairs are identified. 
In Fig.~\ref{theta}, the open circles denote neighbours which have $K_s$ magnitudes, whereas the green crosses denote neighbours which do not have $K_s$ magnitude. 
As is evidenced in Fig.~\ref{theta}, there are a large number ($\sim$43\%) of such neighbours which do not have Ks band magnitudes. This has the consequence of reducing the pool of neighbours from which the major pairs are identified in the Ks filter leading to an underestimation of the major pair fraction in the Ks filter.
Most of the neighbours which do not have measurable $K_s$ magnitude are rather faint, with the green crosses clustering towards the faint end. 
Although the $5h_{100}^{-1}\kpc$ inner cutoff guards against blending in the Ks filter at lower redshifts, in the highest redshift bin, the angular size corresponding to $5h_{100}^{-1}\kpc$ is small enough to allow legitimate neighbours to be identified in the $z$ band within $\leq 1.0\arcsec$ of the host galaxy, where blending in the $K_s$ filter can be a serious concern.   
On further investigation, the bias is indeed found to be more severe in case of close neighbours, with  $\sim$54\% of the neighbours within 1\arcsec~ of the main galaxy not having measured $K_s$ magnitude. Furthermore, even out of the neighbours within 1\arcsec~ of the main galaxy which have a $K_s$ magnitude ascribed to them, $\sim$17\% are actually blended with the host galaxy.
Incompleteness of the $K_s$ band catalog due to detection limit and blending is therefore the main reason for missing out $K_s$ magnitudes for many of the neighbours. This implies that the major merger fraction that we derive using the $K_s$ filter is a {\it{lower limit}}. In addition, this effect is more severe at higher redshifts, leading to an underestimation of the {\it{rate}} at which the merger fraction evolves with redshift. 

In the $z$ band on the other hand, we have verified using the $z$ band number counts in the GOODS-S field, that the photometric catalog is complete down to the faintest magnitude expected of a neighbour for the faintest primary galaxy ($M_J =-19.5$), even at the highest redshift ($\sim$1.2).

%\plot{923844}{An example of a pair of galaxies at redshift 0.273 in the HST/ACS F850LP filter. The image is 18 arcsec in size}{1}

%\plot{240864_Ks}{The same pair of galaxies as in Fig. 1 but in Ks filter. The image is again 18 arcsec in size}{2}

%\begin{eqnarray}
%\log A_e &=& (1.34 \pm 0.28) \log \sigma   {} \nonumber \\
	%& & +{} (0.25 \pm 0.05) \msbe  + \gamma.
%\labequn{sfp}
%\end{eqnarray}

\section{Visible versus near-IR identification of pairs}
\label{bias}

\begin{figure}[h] \centering
   \includegraphics[angle=0, width=0.5\textwidth]{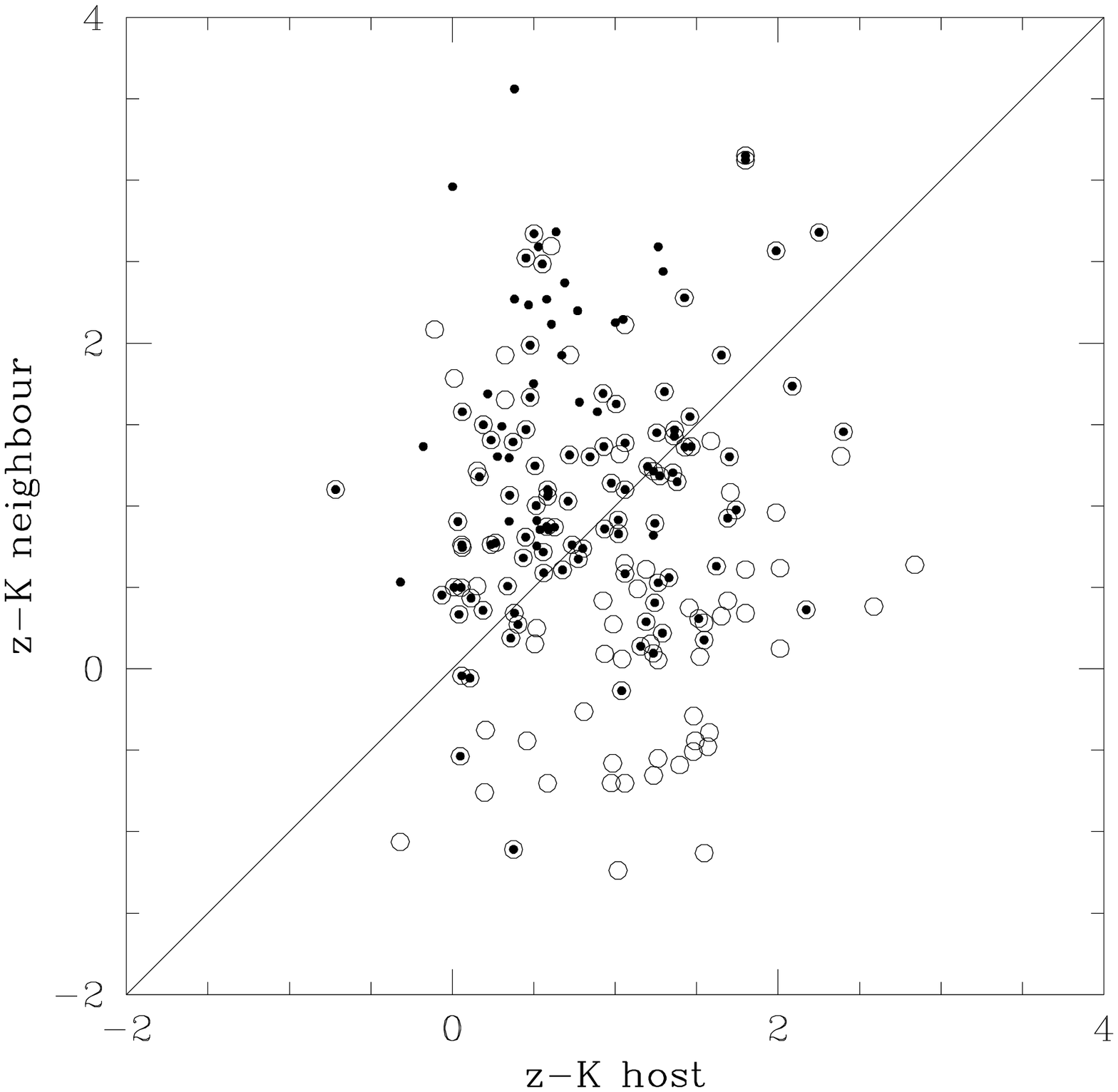}%
   \includegraphics[angle=0, width=0.5\textwidth]{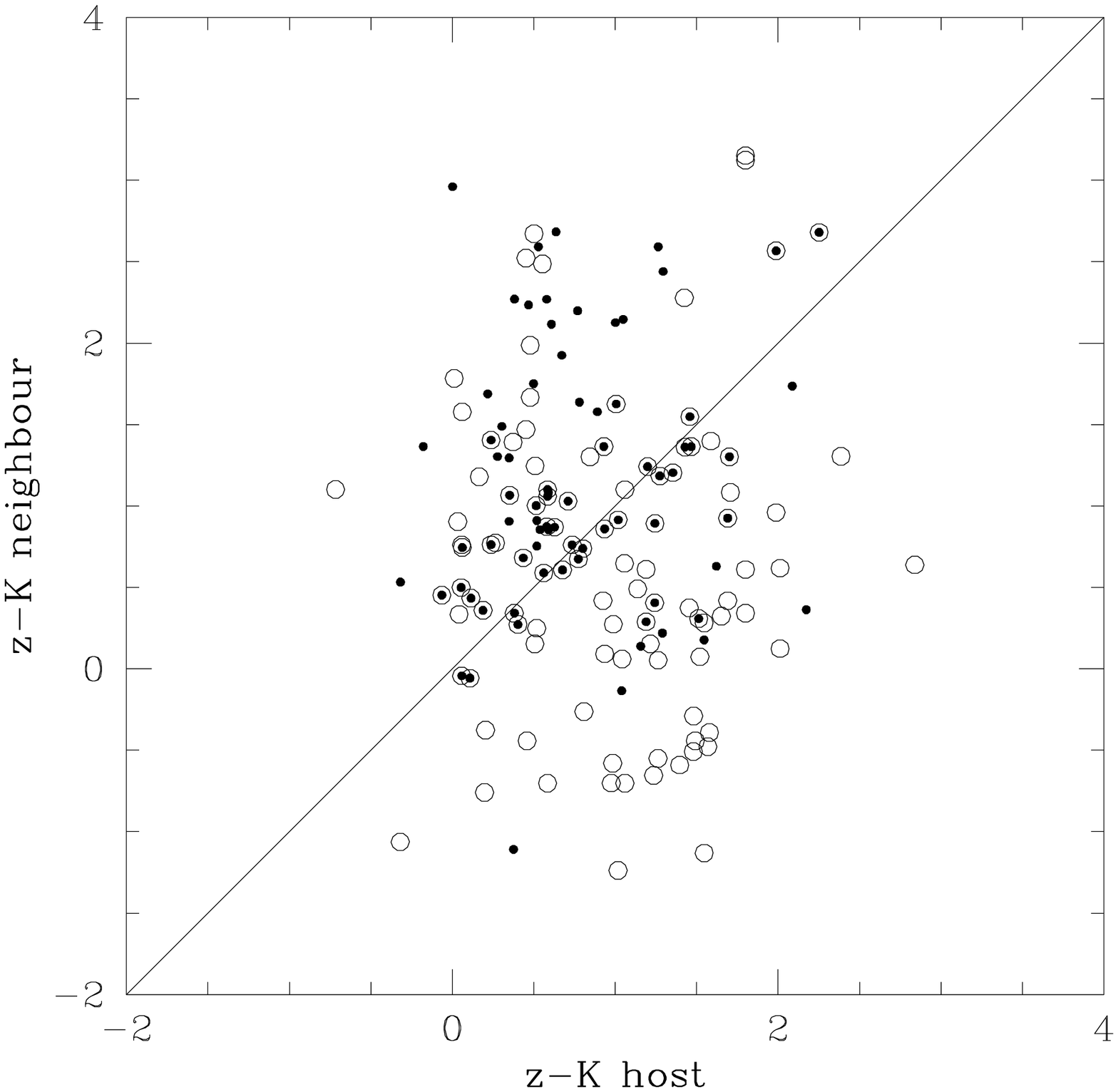}
   \caption{The left figure shows $z-Ks$ color of the primary galaxy Vs that of the {\it{major}} neighbour ($-1.5 \leq \delta m \leq 1.5$) as described in Section~\ref{bias}. The open circles and dots represent neighbours identified in the $z$ filter and in the $K_s$ filter, respectively. Not surprisingly, in the $K_s$ filter we preferentially select neighbours which are redder than the primary galaxy, while a selection in the $z$ filter selects bluer neighbours. The figure on the right is similar to that on the left except that only those major neighbours that are {\it{fainter}} than the primary galaxies are plotted.}
   \label{color}
\end{figure}

As reported by Bundy et al.~\cite{bundy2004}, the selection of pairs in the visible bands may be affected by star formation, i.e. some low mass companions may be brightened in visible, leading to an overestimate of the actual fraction of major pairs. In order to check this assertion, we compare the $z-Ks$ color of the primary galaxy to that of the neighbours which were identified in Section~\ref{pairs}. This is shown in Fig.~\ref{color} (left). Not surprisingly, in the $K_s$ filter we preferentially select neighbours which are redder than the primary galaxy, while a selection in the $z$ filter selects bluer neighbours. In the cases where the $z-Ks$ color of the neighbour is roughly similar to that of the primary galaxy, they are picked out as major neighbours in both the filters.
It must be noted however that neighbours are equally distributed on both sides of the $(z-Ks)_{host}$=$(z-Ks)_{neighbour}$ line. 
%In the cases where the $z-Ks$ color of the neighbour is roughly similar to that of the primary galaxy, they are picked out as major neighbours in both the filters as expected. However, it is to be noted that neighbours can be both redder as well as bluer than the primary galaxy. 
In particular, we do not find evidence for the claim made by Bundy et al.~\cite{bundy2004} that satellite galaxies tend to be bluer than the primary galaxies. In our work the secondary galaxies can be brighter than the primary galaxies, whereas the Bundy et al.~\cite{bundy2004} companions were always fainter than the main galaxies. In order to avoid any incompatibility in comparing our results with those of Bundy et al., we replotted Fig.~\ref{color} (right) using only those neighbours which are fainter than the host galaxy. We again find that the neighbours are equally distributed on both sides of the $(z-Ks)_{host}$=$(z-Ks)_{neighbour}$ line.
This shows that intrinsically the neighbour can be both redder as well as bluer compared to the host galaxy.

%\begin{figure}[h!]
%\includegraphics[height=4.5cm,clip]{917636_compress.ps}%
%\includegraphics[height=4.5cm,clip]{140247_Ks_compress.ps}
%\caption{An example of a pair of galaxies, with the host galaxy having a redshift of 0.35 in the HST/ACS $z$ filter (left) and in $K_s$ filter (right). North is up, East is to the left. The circle in each figure has radius $20 h_{100}^{-1}$\kpc. The neighbour due south-west of the host galaxy is too faint in the $z$ filter to qualify as a major neighbour, but brightens up in the $K_s$ filter and is classified as a major neighbour.}
%\label{bias}
%\end{figure}

%
%\plot{dpps}{Variation of $\delta_{\rm PP}$ with $\log \sigma$, the 
%correlation coefficient is -0.55 at a significance level of 99.94\%}{1}
% 
%
%\subsection{$\delta_{\rm PP} - \log \sigma$ Correlation}
%\labsubsecn{del}
%
\section{The pair statistics and the merger rate evolution with redshift}
\label{evolution}

Table~\ref{table1} lists the primary galaxy/neighbour statistics from our work, segregated in three redshift bins for each of the two filters, $z$ and $K_s$. We have defined the major merger fraction (i.e. the fraction of galaxies likely to merge) as

\begin{eqnarray}
f(z) = \frac{N_{pairs} + N_{pairs}(conf.) - N_{proj.}}{N_{host}} \times 0.5 
\label{frac}
\end{eqnarray}

where $N_{host}$ is the number of primary galaxies in each redshift bin, $N_{pairs}$ is the number of galaxies existing in major pairs within {\it{projected}} $5h_{100}^{-1}\kpc\leq r \leq 20h_{100}^{-1}$\kpc~ of the primary galaxies in either of the two filters, $N_{pairs}(conf.)$ is the number of galaxies spectroscopically confirmed to be existing in {\it{major}} pairs and $N_{proj.}$ is the number of major neighbours expected to be found within $5h_{100}^{-1}\kpc\leq r \leq 20h_{100}^{-1}$\kpc~ of the primary galaxies by pure chance coincidence, estimated statistically using number counts of sources in either of the two filters. The factor $0.5$ is the fraction of close pairs that are likely to merge as estimated by Patton et al.~\cite{patton1997} for z=0 galaxies. This factor is possibly a function of redshift (Lavery et al. 2004), but the form of this dependence is not well established, and is one of the more uncertain factors in our study. While we have corrected our pair fraction for the foreground/background contamination, we do not have at our disposal measurements of relative velocity for the two galaxies found in close pairs in most cases. For some of the pairs where we do have spectroscopic redshifts for both the members, we have put a cutoff on delta velocity $\Delta v$ $\leq$ 1800 km/sec for identifying true pairs. As explained in Section~\ref{test}, this $\Delta v$ $\leq$ 1800 km/sec is to be treated as an upper limit owing to the errorbars on the spectroscopic redshifts used and the true delta velocity is liable to be much smaller. 
This justifies our use of the factor 0.5 in Equ.~\ref{frac} following Patton et al.~\cite{patton2002}, who have studied z$<$ 0.55 galaxies in pairs with dynamical measurements, and find that roughly half of those with $\Delta v$ $<$ 500 km/s, are likely to be merging systems.
% justifying our use of the factor $0.5$.
%It is supposed to evolve with the redshift (see Carlberg et al.~\cite{carlberg2000}) assuming an evolution of the galaxy-galaxy correlation length. The uncertainty of such an evolution is probably one of the largest uncertainty of our study. Because Carlberg et al.~\cite{carlberg2000} found -1$<$x$<$1, and because the correlation length evolves quite slowly in the considered redshift range, we choose x=0 consistently with Carlberg et al.

\begin{figure*}[t]
\includegraphics[height=8.6cm,clip]{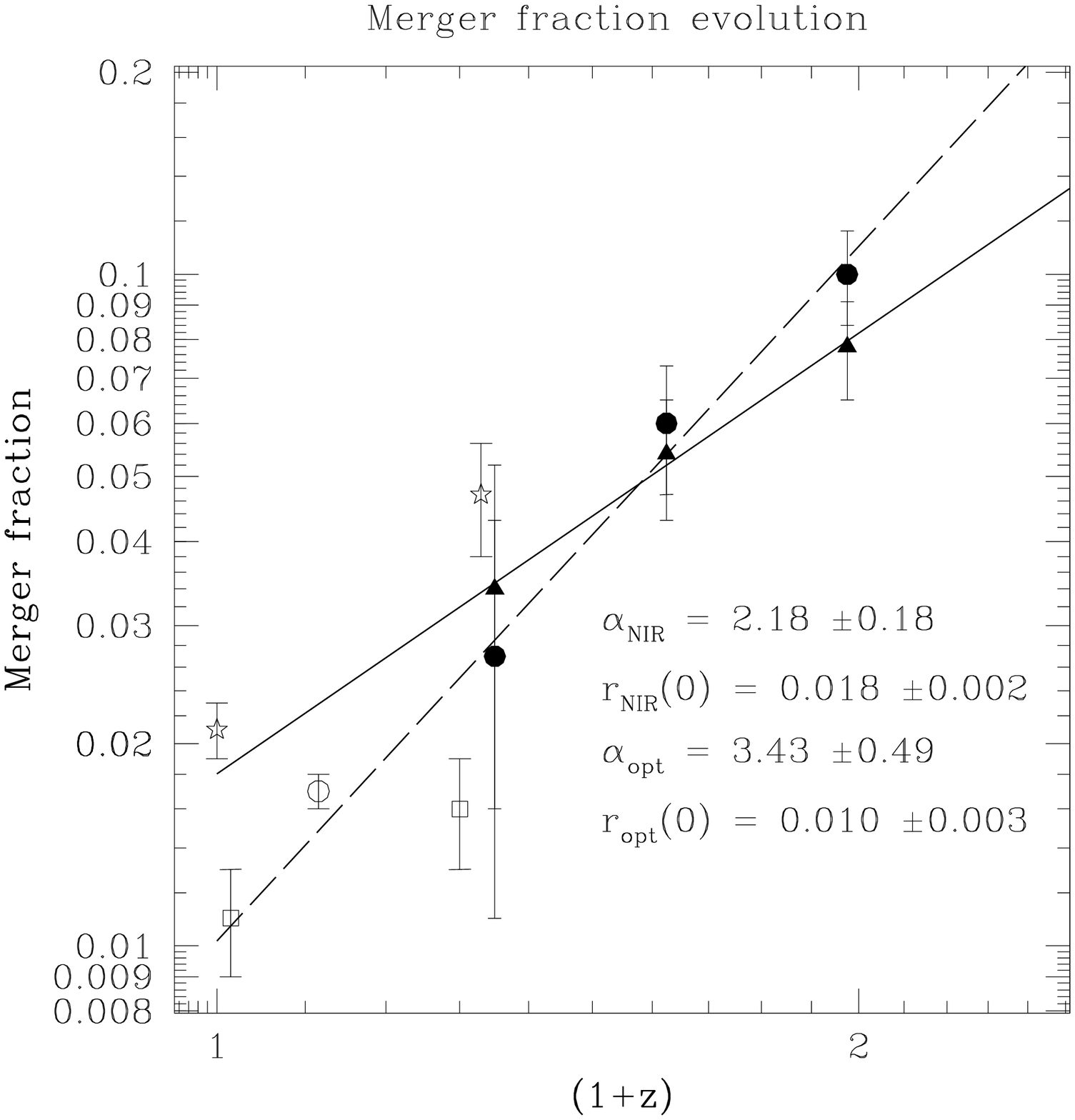}%
\includegraphics[height=8.6cm,clip]{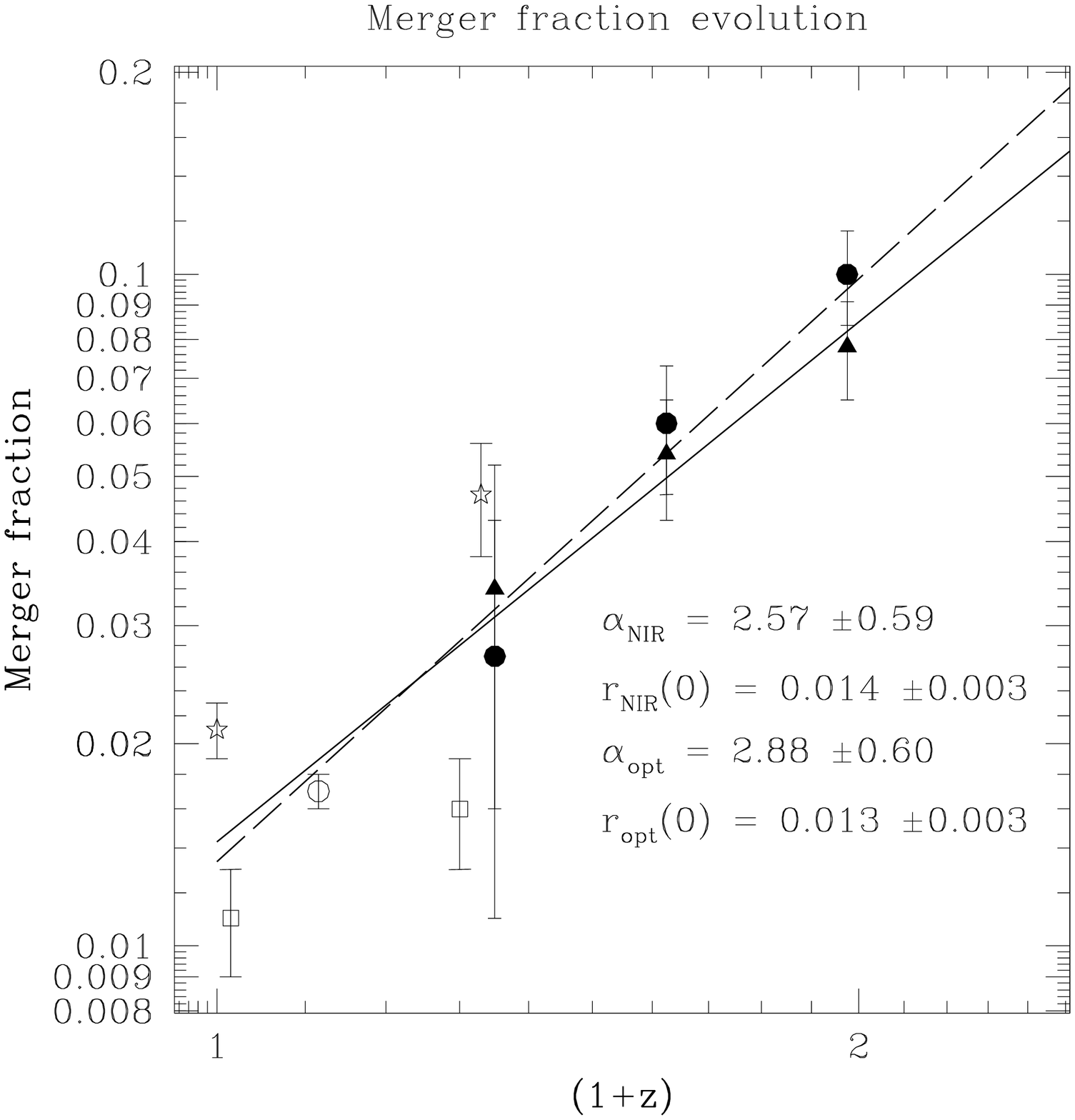}
\caption{The merger fraction evolution with redshift in $z$ filter (filled circles) and $K_s$ filter (filled triangles). The open stars are from Patton et al.~\cite{patton1997}, the open squares are from Patton et al.~\cite{patton2000},~\cite{patton2002} and the open circle is from the Millennium Galaxy Catalogue (De Propris et al. 2005). In the left figure, the dashed line is the best fit curve of the form $f(z)=f(0)\times(1+z)^{\alpha}$ to our $z$ filter datapoints from this work, whereas the solid line is the best fit curve to our $K_s$ filter datapoints from this work. The lower redshift points from literature are just shown on this figure for comparison. In the right figure, the lower redshift points from literature are included in obtaining the fit so that the dashed line is the best fit curve to our $z$ filter datapoints from this work plus all the five lower redshift points from literature, and the solid line is the best fit curve to our $K_s$ filter datapoints plus the five lower redshift points.}
\label{z_Ks_evolution}
\end{figure*}

The evolution in the merger fraction f(z) is plotted in Fig.~\ref{z_Ks_evolution}. In the left panel, the dashed line is the best fit curve of the form $f(z)=f(0)\times(1+z)^{\alpha}$ to our $z$ filter datapoints (filled circles) from this work, whereas the solid line is the best fit curve to our $K_s$ filter datapoints (filled triangles). The lower redshift points taken from literature use sample selection criterion which are inconsistent with our work and are shown only for comparison. The open stars are from Patton et al.~\cite{patton1997}, the open squares from Patton et al.~\cite{patton2000},~\cite{patton2002} and the open circle is from the Millennium Galaxy Catalogue (De Propris et al. 2005). All these works with the exception of Patton et al.~\cite{patton1997} choose their primary sample in the rest frame B, $-21\leq M_B \leq-18$. 
The difference in the slopes obtained from the two filters $z$ and $K_s$ can be traced back to the photometric incompleteness and blending issues in the $K_s$ filter as explained in detail in Section~\ref{blending}. 

The merger fraction shows strong evolution with redshift in both the $z$ as well as the $K_s$ band. However, our derived evolution for merger fraction suffers acutely from a lack of consistent datapoints at redshift$\sim$0.0. The GOODS dataset is based on a relatively small solid angle area (as expected for a deep survey), and is not ideal for deriving the merger fraction at redshifts lower that $\sim$0.2, given the small sample size in this redshift range. On the other hand, the lower redshift estimates of merger fraction that we have available from the literature are inconsistent with our sample selection criterion and are hence not comparable.
 It is for this reason that we utilize the derived parameters from the fit that we have obtained using only our dataset (Fig.~\ref{z_Ks_evolution}(a)), even though it comes at the cost of larger error in terms of the merger rate that we finally derive. 

To get an idea of how much our result is likely to change if we had a pivotal datapoint at redshift $\sim$0.0,
we have refitted the power law curves, this time including the lower redshift points from literature in obtaining the fit. This is shown in Fig.~\ref{z_Ks_evolution} right panel, where the dashed line is the best fit curve to our $z$ filter datapoints (filled circles) {\it{plus}} all the five lower redshift points from literature. Similarly, the solid line is the best fit curve to our $K_s$ filter datapoints (filled triangle) plus the five lower redshift points. The tacit assumption here is that the hypothetical redshift $\sim$0.0 datapoint is some sort of an average of all the low redshift datapoints reported in literature. We notice that even in this case, the merger fraction shows strong evolution with redshift in both the $z$ as well as the $K_s$ band, with the rate of evolution in the $K_s$ filter being slightly shallower on account of the incompleteness and blending issues described earlier. Since this fit was done solely with the purpose of giving an idea to the reader of the sensitivity of our derived results to the presence/absence of a redshift$\sim$0.0 datapoint, the results from this fit are not used anywhere in our work.

%This is done solely for the purpose of giving an idea to the reader of how much our results are likely to change if we had a pivotal datapoint at redshift $\sim$0.0, and the results from this fit are not used anywhere in our work. 

%In Fig.~\ref{z_Ks_evolution} right panel, the lower redshift points from literature are included in obtaining the fit so that the dashed line is the best fit curve to our $z$ filter datapoints (filled circles) {\it{plus}} all the five lower redshift points from literature. Similarly, the solid line is the best fit curve to our $K_s$ filter datapoints (filled triangle) plus the 5 lower redshift points. This is done solely for the purpose of giving an idea to the reader of how much our result is likely to change if we had a pivotal datapoint at redshift $\sim$0.0, and the results from this fit are not used anywhere in our work. 

We used a prescription similar to that of Lin et al.~\cite{lin2004} to convert the merger fraction (fraction of galaxies likely to merge) into a comoving merger rate (i.e. number of merger {\it{events}}/$Mpc^{3}$/Gyr)

\begin{eqnarray}
N_{mg}(z) = 0.5n(z)f(z)\tau^{-1}
\end{eqnarray}

where $n(z)$ is the comoving number density of galaxies (obtained by integrating the Pozzetti et al.~\cite{pozzetti2003} luminosity function over the luminosity range of interest), $f(z)$ is the merger fraction derived above, the factor $0.5$ converts the number of merging galaxies into number of merger events and $\tau$ is the merger timescale (assumed to be 0.5\gyr). This yielded a merger rate of $2.08\times10^{-4} \mpc^{-3}\gyr^{-1}$ at the current epoch, which evolves by a factor $\sim5.6$ to $1.16\times10^{-3} \mpc^{-3}\gyr^{-1}$ at $z=1.2$ using $K_s$ data. For the $z$ band data, these numbers are  $1.16\times10^{-4} \mpc^{-3}\gyr^{-1}$ at the current epoch, which evolves by a factor $\sim15$ to $1.73\times10^{-3} \mpc^{-3}\gyr^{-1}$ at $z=1.2$. Integrating over the timescale from $z=1.2$ to the current epoch, we find that $41\%\pm3.5\% (45\%\pm13\%$ for the z filter) of galaxies with $M_J\leq-19.5$ have undergone a major merger in the last $\sim8 \gyr$. The merger timescale is generally estimated from simulations to range from 0.1 to 1\gyr (see Hernquist and Mihos 1995), which brings an additional uncertainty to our result. For example if it were assumed to be 0.35\gyr (Carlberg et al. 2000), it would lead to 59\% as being the fraction of galaxies having experienced a major merger in the last $\sim8 \gyr$.

%\begin{figure}[h] \centering
 %  \includegraphics[angle=0, width=0.4\textwidth]{z_Ks_evolution.err2.ps}
  % \caption{The merger fraction evolution with redshift in $z$ filter (open circles) and $K_s$ filter (filled triangles). The filled squares are from Patton et al.~\cite{patton1997}. The dashed line is the best fit curve of the form $f(z)=f(0)\times(1+z)^{\alpha}$ to the $z$ filter datapoints plus the lower redshift points from Patton et al.~\cite{patton1997}, whereas the solid line is the best fit curve to the $K_s$ filter datapoints plus the lower redshift points from Patton et al.~\cite{patton1997} } 
   %\label{z_Ks_evolution}
%\end{figure}

\subsection{The effect of the Large Scale Structure in the CDFS}
As is well reported in literature (Le Fevre et al. 2004, Ravikumar et al. 2007), there are several large scale structures in the CDFS which show up as spikes in the redshift distribution obtained from various redshift surveys. The most prominent large scale structure is at a redshift of 0.735. It represents a factor $\sim$10 overdensity in terms of galaxy number density. This has the potential of significantly altering the results that we have obtained for merger fraction in the redshift bin $0.5 \leq z \leq 0.75$. The detailed effect of such large scale structures on the cosmological relevance of the GOODS south field have been considered by Ravikumar et al.~\cite{ravi2007}. 

In order to check the robustness of our derived merger fraction, we recalculated the merger fraction in the redshift bin $0.5 \leq z \leq 0.75$ by excluding the sources embedded in the large scale structure at z=0.735 (defined as objects within $\delta v \leq 1500 \kms$ of z=0.735). This resulted in the exclusion of 70 sources from the bin $0.5 \leq z \leq 0.75$ out of which 11 were major pairs (in the Ks band). The merger fraction dropped from 5.48\% to 5.34\% which is well within the reported errorbars. The derived merger rate is therefore not affected by the presence of the large scale structure.

\section{Discussion}
\label{discuss}
We have derived the major merger rates of galaxies in the redshift range $0.2 \leq z \leq 1.2$ using pair counting in both optical and near-IR bands. This work provides a robust estimate of the major merger rate upto redshift $\sim$1.2 using a representative sample of near-IR selected galaxies, which are well within the photometric completeness of the source catalogs. We find steep evolution with redshift, with the merger fraction $\propto (1+z)^{3.43\pm0.49}$ for optically selected pairs, and $\propto (1+z)^{2.18\pm0.18}$ for pairs selected in the near-IR. The difference in the slopes obtained from the two filters $z$ and $K_s$ can be traced back to the photometric incompleteness and blending issues in the $K_s$ filter.
%Assuming a redshift dependence of the form $\propto (1+z)^{\alpha}$ for the major merger fraction, we find a steep evolution with redshift in {\it{both}} optical as well as near-IR selected samples, with $\alpha_{optical}=2.44\pm0.39$ \& $\alpha_{near-IR}=2.07\pm0.74$. 
We find that the major merger rate evolves by a factor $\sim5.6$ from $2.08\times10^{-4} \mpc^{-3}\gyr^{-1}$ at the current epoch to $1.16\times10^{-3} \mpc^{-3}\gyr^{-1}$ at $z=1.2$. This implies that between 41\% and 59\% of all galaxies with $M_J\leq-19.5$ have undergone a major merger in the last $\sim8 \gyr$, assuming that the average timescale for a pair to merge is in the range 0.5 to 0.35\gyr.

%This implies that $58\%\pm17\%$ of all galaxies with $M_J\leq-19.5$ have undergone a major merger in the last $\sim8 \gyr$, assuming that the average timescale for a pair to merge is 0.5\gyr. 
%

Our result is in agreement with the recently published results by the COSMOS team (Kartaltepe et al. 2007) who report a power law index of $3.1\pm0.1$, as well as other works such as Le fevre et al.~\cite{lefevre2000}, Carlberg et al.~\cite{carlberg1994}, Patton et al.~\cite{patton2002} (see Table 2 in Kartaltepe et al. 2007). Reports of a significantly lower merger rate (Lin et al. 2004, Carlberg et al. 2000), might be due to the fact that the investigators have used a luminosity evolution model for sample selection. Indeed such a correction is not significant in our work since we have selected our sample in the rest frame near-IR.

The disagreement in our $K_s$ derived merger fraction evolution with that reported by Bundy et al.~\cite{bundy2004} might be due to their relatively small statistics (our sample size is $\sim4$ times larger than theirs), and that they did not account for neighbours that are {\it{brighter}} than the primary galaxy. Our study implies that the star formation enhancement of the companion does not affect severely the calculation of the merger rate. 

We notice that at least some of the discrepancies between results derived in this paper and those of other workers in the field can be attributed to different definitions of major mergers. The methodology of Lotz et al.~\cite{lotz2006} probes a merger at a different stage in the merging process (ongoing merger), whereas our definition of major merger is sensitive to pairs of galaxies in an early stage of merger (upcoming mergers) so that two distinct members are easily distinguishable. Studies with 3D spectroscopy are certainly better suited to identify on-going or post-merger stages. Such studies (Flores et al. 2006, Yang et al. 2007) have found a frequent presence of complex kinematics at z=0.6 (26\% of $M_J$$<$-20.3 galaxies), which if related to mergers, is 4 to 5 times higher than paired/on going mergers found in studies like this one (see also Lotz et al. 2006). Indeed Hammer et al.~\cite{hammer2007} noticed that all studies of merger rate agree to find 5\%$\pm$$1\%$ for the merger fraction at z=0.6. If complex kinematics are related to mergers, it may imply a long duration (1-2 Gyr) for the remnant phase during which gaseous velocity fields are severely perturbed or chaotic (using a simple ergodic argument).
%Such a high fraction of on-going/post mergers is 4 times larger than what is found in this study ($\sim$1\%
 %of z=0.6 galaxies being in pairs which are likely mergers), which may imply that the merging remnant phase is longer than the encountering phase using the simple ergodic argument (eg. Hammer et al.~\cite{hammer2007}). 
 In the near future we will combine 3D spectroscopy of paired galaxies to identify the different time-scales of the merging process which might shape the galaxies as we observe them today.
%A simple argument using the ergodic principle should convince the reader that the fraction of objects identified as major mergers using the two definitions will be different, and will reflect the underlying amount of time a pair spends in each of the two stages of the merging process. Hence our result is not directly comparable to that of Lotz et al.~\cite{lotz2006}, and the two methods may be considered to be complimentary.   
%

%We also quantified the differences between the properties of paired and isolated galaxies. We find that paired galaxies are slightly bluer with a median $B-z=1.57$ compared to $B-z=2.08$ for isolated galaxies. The median SFR for paired galaxies with detectable 24$\mu$m fluxes is found to be marginally higher at $\sim 32M_{\odot}/yr$ compared to $\sim 23M_{\odot}/yr$ for isolated galaxies. This might be indicative of enhanced star formation due to the interaction between the members of the pair.
%

%

Finally we notice that although Toomre type (Toomre \& Toomre 1972) tidal features are seen in close pairs of galaxies in the lower redshift bins, such features are conspicuously absent in candidate pairs at higher redshifts. This is likely due to the $(1+z)^{-4}$ surface brightness dimming, which quickly pushes the low surface brightness tidal features below the detection limits at higher redshifts.

\acknowledgments We thank the Centre Franco-Indien pour la Promotion de la Recherche Avancee (CEFIPRA) for financial assistance under project number 2804-1. A.R. would like to thank CSIR for PhD. funding.

\clearpage

\begin{deluxetable}{lccccccccc}

\tablecaption{Pair Statistics}
%\caption{Pair Statistics}
\rotate

\tablewidth{0pt}
\tablehead{\colhead{$z$} & \colhead{$N_{host}$} & \colhead{$N_{pairs}^{F850LP}$}& \colhead{$N_{pairs}^{F850LP}(conf.)$}  & \colhead{$N_{proj.}^{F850LP}$}   & \colhead{$N_{pairs}^{Ks}$}& \colhead{$N_{pairs}^{Ks}(conf.)$}  & \colhead{$N_{proj.}^{Ks}$} & \colhead{f(z)(F850LP)} & \colhead{f(z)(Ks)}}
\startdata
%$z$ & $N_{host}$ & $N_{pairs}(F850LP)$ & $N_{proj.}(F850LP)$   & $N_{pairs}(Ks)$  & $N_{proj.}(Ks)$ & Merger Fraction(F850LP) & Merger Fraction(Ks)\\
% & & (F850LP) & (F850LP) & (Ks) & (Ks) & (F850LP) & (Ks) \\

0.20--0.50  &  61 & 4 & 0 &  0.70 & 5 & 0 & 0.81 & 2.70 $\pm$ 1.67  & 3.43 $\pm$ 1.87  \\
0.50--0.75  &  294 & 48 & 7  & 19.56 &  39 & 3  & 9.76 & 6.03 $\pm$ 1.30  & 5.48 $\pm$ 1.14  \\
0.75--1.20 &  340 & 98 & 13 & 42.94 & 62 & 10  & 18.80 & 10.01 $\pm$ 1.64 &  7.82 $\pm$ 1.31 \\
\hline
\hline
TOTAL & 695 & 150 & 20 & 63.20 & 106 & 13 & 29.37 & -- & -- \\
\enddata

\tablecomments{$N_{host}$ is the number of primary galaxies in each redshift bin. $N_{pairs}$ is the number of galaxies existing in major pairs within {\it{projected}} $5h_{100}^{-1}\kpc\leq r \leq 20h_{100}^{-1}$\kpc~ of the primary galaxies in each of the two filters. $N_{pairs}(conf.)$ is the number of galaxies spectroscopically confirmed to be existing in {\it{major}} pairs. $N_{proj.}$ is the number of major neighbours expected to be found within $5h_{100}^{-1}\kpc\leq r \leq 20h_{100}^{-1}$\kpc~ of the primary galaxies by pure chance coincidence, estimated statistically using number counts of sources in each of the two filters. Merger fraction is estimated as $(N_{pairs}+N_{pairs}(conf.)-N_{proj.})/N_{host} \times 0.5$. The errorbars are estimated using Poisson statistics.  \label{table1} }

\end{deluxetable}

\clearpage

\clearpage

\clearpage

\clearpage

\clearpage

\clearpage

\clearpage

\clearpage

\clearpage

%\begin{figure}[h] \centering
 %  \includegraphics[angle=0, width=1.0\textwidth]{f10.eps}
  % \caption{$B-z$ color distribution of paired {\it{host}} galaxies(hashed) compared to that of isolated galaxies.}
   %\label{b_z_color}
%\end{figure}

\end{document}